\documentclass[journal]{IEEEtran}

\usepackage{cite}
\usepackage{textcomp}
\usepackage{amsmath, amssymb, amsfonts}
\usepackage{accents}
\usepackage{algorithm}
\usepackage{algorithmic}
\usepackage{color}
\usepackage{graphicx}
\usepackage{enumerate}
\usepackage{booktabs}
\usepackage{subcaption}
\usepackage{mathrsfs}
\usepackage{mathtools}
\usepackage{dsfont}
\usepackage{relsize}

\DeclareMathOperator{\tr}{tr}
\DeclareMathOperator{\minimize}{minimize}

\DeclareMathOperator{\diag}{diag}

\DeclareMathOperator{\E}{\mathsf{E}}
\DeclareMathOperator{\Cov}{\mathsf{cov}}

\DeclareMathOperator{\ProbM}{\mathsf{P}}
\DeclareMathOperator{\Prob}{\mathsf{p}}

\newtheorem{definition}{Definition}
\newtheorem{lemma}{Lemma}

\newtheorem{theorem}{Theorem}

\newtheorem{corollary}{Corollary}
\newtheorem{remark}{Remark}

\makeatletter
\let\NAT@parse\undefined
\makeatother
\usepackage{hyperref}  
\hypersetup{
    colorlinks= true,
    citecolor= black,
    linkcolor= black, 
    pdfborder= {0 0 0},
}

\begin{document}

\title{State Estimation over Broadcast and Multi-Access Channels in an Unreliable Regime}
\author{
Touraj Soleymani, \emph{IEEE}, \emph{Member}, and Deniz G\"{u}nd\"{u}z, \emph{IEEE}, \emph{Fellow}
\thanks{Touraj Soleymani and Deniz G\"{u}nd\"{u}z are with the Department of Electrical and Electronic Engineering, Imperial College London, London SW7 2AZ, United Kingdom (e-mails: {\tt\small touraj@imperial.ac.uk}, {\tt\small d.gunduz@imperial.ac.uk}).}%
}
\maketitle

\begin{abstract}
This article examines the problem of state estimation over multi-terminal channels in an unreliable regime. More specifically, we consider two canonical settings. In the first setting, measurements of a common stochastic source need to be transmitted to two distinct remote monitors over a packet-erasure broadcast channel. In the second setting, measurements of two distinct stochastic sources need to be transmitted to a common remote monitor over a packet-erasure multi-access channel. For these networked systems, we uncover the fundamental performance limits in the sense of a causal tradeoff between the estimation error and the communication cost by identifying optimal encoding and decoding strategies. In the course of our analysis, we introduce two novel semantic metrics that play essential roles in state estimation over broadcast and multi-access channels. The first metric arising in the context of broadcast channels is the \emph{dissemination value of information}, which quantifies the valuation of provisioning a piece of information to multiple receivers simultaneously. The second metric arising in the context of multi-access channels is the \emph{prioritization value of information}, which quantifies the valuation of provisioning a piece of information chosen from one out of multiple transmitters. Our findings certify that the optimal encoding and decoding strategies hinge on these semantic metrics.
\end{abstract}

\vspace{4mm}

\section{Introduction}
\IEEEPARstart{T}{he estimation} of the state of a dynamical system from its measurements is a crucial problem, manifesting across various domains, as the state provides a mathematical representation of the system's behavior at each time, which can be used as a basis for decision making~\cite{kalman1960, kalman1963}. Nevertheless, in many real-world scenarios such as space exploration, sensors that acquire measurements and nodes that compute state estimates are physically separated and connected through communication channels that are subject to various constraints and impairments. The presence of these communication channels can severely affect the system performance and can pose major challenges to the design procedure. Many works have endeavored to address such challenges~\cite{imer2010, rabi2012, lipsa2011, lipsa2009optimal, voi, voi2, molin2017, chakravorty2016, chak2016loss, chak2017loss, guo2021-IT, guo2021-TAC, sun2019, vasconcelos2018opt, zhang2021dis, vasconcelos2021data, vasconcelos2019obs, sun2017, kadota2018scheduling, ceran2019average, ceran2021reinf, kosta2019age, bhat2021minimization, he2017optimal, kushner, meier1967, mywodespaper, soleymani2016-cdc, leong2017, molin2019, witsenhausen1979, walrand1983, borkar2001, yuksel2012, tanaka2016, khina2018t, sinopoli, wu2017, quevedo2013, schenato2008delay, gupta2009d}.

\begin{figure}[t]
\centering
  \includegraphics[width=.98\linewidth]{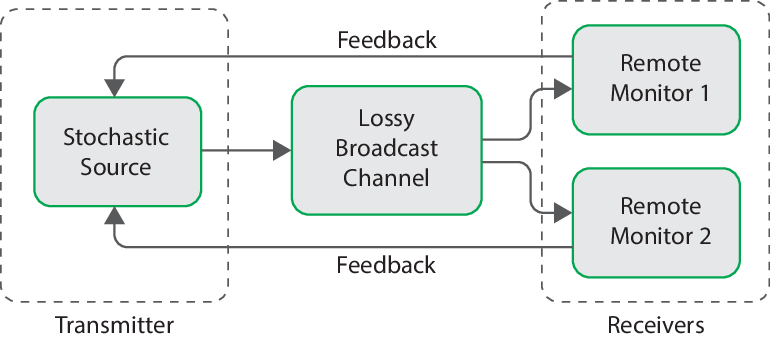}
  \caption{State estimation of a stochastic source over a lossy broadcast channel.}
  \label{fig:schematic-bc}
\end{figure}

\begin{figure}[t]
\centering
  \includegraphics[width=.98\linewidth]{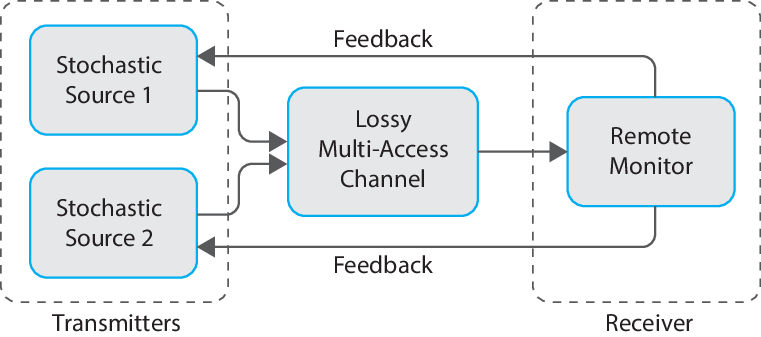}
  \caption{State estimation of stochastic sources over a lossy multi-access channel.}
  \label{fig:schematic-mac}
\end{figure}

In the present article, we examine the problem of state estimation over multi-terminal channels in an unreliable regime. More specifically, we consider two canonical settings. In the first setting, depicted in Fig.~\ref{fig:schematic-bc}, measurements of a common stochastic source need to be transmitted over a lossy broadcast channel to two distinct remote monitors for state estimation. In the second setting, depicted in Fig.~\ref{fig:schematic-mac}, measurements of two distinct stochastic sources need to be transmitted over a lossy multi-access channel to a common remote monitor for the same purpose. The rationale for analyzing these networked systems is to comprehend the above-mentioned problem in the simplest forms of multi-terminal communications~\cite{cover1972, gallager1985}, which can serve as a foundation for development of more complex networks. For these networked systems, which are supposed to operate in real-time, we are interested in uncovering the fundamental performance limits in the sense of a causal tradeoff between the estimation error and the communication cost, a quest through which we hope to gain deeper insights into the principles of \emph{semantic communications}~\cite{uysal2022semantic, gunduz2022semantic}, where the goal is to exchange only the most significant part of data by taking into account its context. These performance limits can clearly be obtained by identifying optimal encoding and decoding strategies, which will be the central focus of our~analysis.

\subsection{Related Work}
A close look at existing studies in the literature on the design of encoding and decoding policies for real-time networked systems reveals four distinct classes. The first class of studies examines a causal tradeoff between the mean square error (MSE) and the transmission frequency~\cite{imer2010, rabi2012, lipsa2011, lipsa2009optimal, voi, voi2, molin2017, chakravorty2016, chak2016loss, chak2017loss, guo2021-IT, guo2021-TAC, sun2019, vasconcelos2018opt, zhang2021dis, vasconcelos2021data, vasconcelos2019obs}. We refer to the problem in these studies, where the encoding policy is an observation-based scheduling policy and the decoding policy is an estimation policy, as \emph{remote estimation}. In this class, for instance, remote estimation of a multi-dimensional partially observable Gauss--Markov process over an ideal channel was studied in \cite{voi, voi2}, remote estimation of a scalar Gauss--Markov process over an ideal channel and an independent and identically distributed~(i.i.d.) packet-erasure channel in \cite{lipsa2011, lipsa2009optimal}, remote estimation of a scalar Markov process with symmetric noise distribution over an ideal channel in \cite{molin2017}, remote estimation of a scalar autoregressive Markov process with symmetric noise distribution over an ideal channel, an i.i.d.~packet-erasure channel, and a Gilbert--Elliott packet-erasure channel in~\cite{chakravorty2016, chak2016loss, chak2017loss}, remote estimation of the scalar Wiener and scalar Ornstein--Uhlenbeck processes over an ideal channel and a fixed-delay channel in~\cite{guo2021-IT,guo2021-TAC}, remote estimation of the scalar Wiener process over a random-delay channel in~\cite{sun2019}, remote estimation of multiple random variables with arbitrary distributions over a collision channel in~\cite{vasconcelos2018opt, zhang2021dis}, remote estimation of multiple random variables with arbitrary distributions over unicast and broadcast channels in~\cite{vasconcelos2021data}, and remote estimation of two Gaussian random variables over a multi-access channel in~\cite{vasconcelos2019obs}. These studies established certain characteristics such as a symmetric, asymmetric, or threshold structure of the optimal scheduling policy with respect to the estimation discrepancy.

The second class of studies investigates a causal tradeoff between an age penalty and the transmission frequency~\cite{sun2017, kadota2018scheduling, ceran2019average, ceran2021reinf, he2017optimal, kosta2019age, bhat2021minimization}. We refer to the problem in these studies, where the encoding policy is an age-based scheduling policy and the decoding policy is often irrelevant, as \emph{status updating}. Note that the age of information is a semantic metric that can indirectly be representative of the quality of state estimation at the receiver. In this class, for example, status updating by a source over a random-delay channel was studied in \cite{sun2017}, where it was shown that the so-called zero-wait scheduling policy cannot be optimal. Status updating by multiple sources over an i.i.d.~packet-erasure broadcast channel was studied in \cite{kadota2018scheduling}, where it was shown that a greedy scheduling policy is optimal when the network is symmetric. Status updating by a source over an i.i.d.~packet-erasure channel and an i.i.d.~packet-erasure broadcast channel with retransmissions is studied in~\cite{ceran2019average, ceran2021reinf}, where the general properties of the optimal scheduling policy were obtained. Moreover, status updating by multiple sources over an ideal multi-access channel, a random-delay multi-access channel, and a fading multi-access channel was studied in \cite{he2017optimal, kosta2019age, bhat2021minimization}, where the complexity of the problem and the structure of the optimal scheduling policy were characterized.

The third class of studies analyzes a causal tradeoff between a variance penalty and the transmission frequency~\cite{kushner, meier1967, mywodespaper, soleymani2016-cdc, leong2017, molin2019}. We refer to the problem in these studies, where the encoding policy is a variance-based scheduling policy and the decoding policy is an estimation policy, as \emph{sensor scheduling}. Note that in sensor scheduling, in contrast to remote estimation, as described above, the realized sensory information is not exploited in the scheduling policy. In this class, previously, sensor scheduling of a Gauss--Markov process over an ideal channel was studied in \cite{meier1967}, where it was proved that the optimal scheduling policy can be obtained by solving a deterministic optimization problem. More recently, sensor scheduling of a Gauss--Markov process over an ideal channel based on estimation entropy was studied in \cite{mywodespaper, soleymani2016-cdc}, where the optimal scheduling policy was derived. Sensor scheduling of a Gauss--Markov process observed by multiple sensors over an~i.i.d.~packet-erasure multi-access channel was studied in~\cite{leong2017}, where it was shown that the optimal scheduling policy  has a threshold-type behavior in switching between different sensors. Furthermore, sensor scheduling of Gauss--Markov processes over an ideal multi-access channel was studied in~\cite{molin2019}, where a suboptimal variance-based scheduling policy was obtained associated with the best linear unbiased estimator at a monitor.

Lastly, the fourth class of studies establishes a causal tradeoff between the MSE~and the bit rate~\cite{witsenhausen1979, walrand1983, borkar2001, yuksel2012, tanaka2016, khina2018t}. We refer to the problem in these studies, where the encoding policy is a compression policy and the decoding policy is an estimation policy, as \emph{sequential coding}. Note that these studies rely on the fact that compressed sensory information is transmitted in a periodic manner. In this class, for instance, sequential coding of a discrete-time $n$th order Markov process over an ideal channel was addressed in~\cite{witsenhausen1979}, sequential coding of a discrete-time finite-state Markov process over a noisy channel with feedback in~\cite{walrand1983}, sequential coding of a partially observable continuous-state Markov process with variable quantization levels in~\cite{borkar2001}, and sequential coding of a partially observable continuous-state Markov process over a multi-access channel in \cite{yuksel2012}. These studies revealed certain properties such as a separate design associated with the optimal compression policy. In addition, sequential coding of a Gauss-Markov process with multiple sensors over an i.i.d.~packet-erasure channel was studied in~\cite{khina2018t}, where the achievable causal rate-distortion region was characterized.

It is worth mentioning that there are also pertinent studies in the literature that have analyzed the severe effects of packet loss on stability of state estimation over communication channels~\cite{sinopoli, wu2017, quevedo2013, schenato2008delay, gupta2009d}. Note that these studies are based on the assumption that sensory information is periodically transmitted by the encoder. In particular, mean-square stability of Kalman filtering over an i.i.d.~packet-erasure channel was studied in~\cite{sinopoli}, peak-covariance stability of Kalman filtering over a Gilbert--Elliott packet-erasure channel in~\cite{wu2017}, and mean-square stability of Kalman filtering over a fading packet-erasure channel with correlated gains in~\cite{quevedo2013}. The results in these studies show that there exists a critical region for the channel condition outside which the underlying networked system can become unstable. Furthermore, various properties of state estimation of a Gauss--Markov process over an i.i.d.~packet-erasure channel were analyzed in~\cite{gupta2009d}, where it was shown that transmitting the minimum mean-square-error (MMSE) state estimate at the encoder at each time leads to the maximal information set for the decoder.

\subsection{Contributions and Outline}
In this article, we aim to identify optimal encoding and decoding policies pertaining to state estimation of partially observable Gauss-Markov processes over time-varying packet-erasure broadcast and multi-access channels. In the course of our analysis, we will introduce two novel semantic metrics that we deem essential for state estimation over multi-terminal channels. The first metric arising in the context of broadcast channels is the ``\emph{dissemination value of information}'', which quantifies the valuation of provisioning a piece of information to multiple receivers simultaneously. The second metric arising in the context of multi-access channels is the ``\emph{prioritization value of information}'', which quantifies the valuation of provisioning a piece of information chosen from one out multiple transmitters. Our findings certify that the optimal encoding and decoding policies hinge on these semantic metrics. Note that the notion of the ``\emph{value of information}'' in the context of feedback control over communication channels was previously introduced in~\cite{touraj-thesis, voi, voi2}. The dissemination value of information and the prioritization value of information should be conceived as natural generalizations of the value of information to multi-terminal~scenarios.

Our study centers on a causal frequency-distortion tradeoff defined between the total MSE~and the total transmission frequency, which sets it inherently apart from~\cite{sun2017, kadota2018scheduling, ceran2019average, ceran2021reinf, kosta2019age, bhat2021minimization, he2017optimal, kushner, meier1967, mywodespaper, soleymani2016-cdc, leong2017, molin2019, witsenhausen1979, walrand1983, yuksel2012, borkar2001, khina2018t, tanaka2016, sinopoli, wu2017, quevedo2013, schenato2008delay, gupta2009d}. Our structural results extend the previous results in~\cite{imer2010, rabi2012, lipsa2011, lipsa2009optimal, voi, voi2, molin2017, chakravorty2016, chak2016loss, chak2017loss, guo2021-IT, guo2021-TAC, sun2019, vasconcelos2018opt, zhang2021dis, vasconcelos2021data, vasconcelos2019obs} to more complex settings. Note that our study differs from~\cite{imer2010, rabi2012, lipsa2011, lipsa2009optimal, voi, voi2, molin2017, chakravorty2016, chak2016loss, chak2017loss, guo2021-IT, guo2021-TAC, sun2019}, which address state estimation over point-to-point channels. We here examine state estimation over broadcast and multi-access channels, which requires taking into account the effects of distinct estimation discrepancies and channel conditions. In addition, our study differs from~\cite{vasconcelos2018opt, zhang2021dis, vasconcelos2021data, vasconcelos2019obs}, which address state estimation of independent random variables, where previous and current decisions do not have any impact on the system performance in the future. We here examine state estimation of Markov processes, which demands sequential decision making.

The rest of the article is structured as follows. The problem of state estimation over a broadcast channel and that over a multi-access channel are formulated separately in Section~\ref{sec:statement}. Our main theoretical results are presented and discussed in Section~\ref{sec:main-results}, followed by the derivation of these results in Section~\ref{sec:der-main-results}. Our numerical results pertaining to satellite communications are provided in Section~\ref{sec:example}. The article is concluded and future research is discussed in Section~\ref{sec:conclusions}.

\subsection{Preliminaries}
Throughout the article, we adopt the following convention. The probability measure of a random variable $x$ is represented by $\mathsf{P}(x)$, its probability density or probability mass function by $\Prob(x)$, and its expected value and covariance by $\E[x]$ and $\Cov[x]$, respectively. Given $(\mathcal{X},\mathcal{B}_{\mathcal{X}})$ and $(\mathcal{Y},\mathcal{B}_{\mathcal{Y}})$ as two measurable spaces, the mapping $\ProbM: \mathcal{B}_{\mathcal{Y}} \times  \mathcal{X} \to [0,1]$ is a Borel measurable stochastic kernel if $\mathcal{A} \mapsto \ProbM( \mathcal{A} | x)$ is a probability measure on $(\mathcal{Y},\mathcal{B}_{\mathcal{Y}})$ for any $x \in \mathcal{X}$, and $x \mapsto \ProbM(\mathcal{A}| x )$ is a Borel measurable function for any $\mathcal{A} \in \mathcal{B}_{\mathcal{Y}}$. The variables $x$ and $y$ are real and non-negative integer, if $x \in \mathbb{R}$ and $y \in \mathbb{N}$, respectively. Given the variables $x,y \in \mathbb{N}$, $x \leq y$, the set $\mathbb{N}_{[x,y]}$ denotes $\{z \in \mathbb{N} | x \leq z \leq y\}$. The sequence of all vectors $x_t$, $t=p,\dots,q$, is represented by $\mathbf{x}_{p:q}$. Given $s \in \mathcal{M}$ for any set $\mathcal{M}$, $\bar{s}$ represents any elements in $\mathcal{M} - \{s\}$. The matrices $X$ and $Y$ are positive definite and positive semi-definite, if $X \succ 0$ and $Y \succeq 0$, respectively. The logical~AND and logical~OR are represented by $\wedge$ and $\vee$, respectively. The indicator function of a subset $\mathcal{A}$ of a set $\mathcal{X}$ is denoted by $\mathds{1}_\mathcal{A}:\mathcal{X} \to \{0,1\}$. A Borel measurable function $f(x)$ is symmetric if $f(x) = f(-x)$, and is radially symmetric if $f(x) = f(\sqrt{x^Tx})$. The symmetric decreasing rearrangement of a Borel measurable function $f(x)$ vanishing at infinity is represented by $f^*(x)$.


\section{Problem Statement}\label{sec:statement}
In this section, we present the detailed mathematical formulation of the causal frequency-distortion tradeoff pertaining to state estimation over broadcast and multi-access channels for the settings depicted in Figs.~\ref{fig:schematic-bc}~and~\ref{fig:schematic-mac}, respectively. To ensure clarity, we formulate each case separately. This formulation will serve as the basis for deriving optimal encoding and decoding policies in the next section.

\subsection{State Estimation over a Broadcast Channel}
Consider a networked system composed of a common stochastic source with an encoder, two distinct remote monitors with decoders, and a broadcast channel that connects the source to the monitors. Let $\mathcal{M}$ be the set of the monitors and $c$ as an index emphasize the fact that the source is common. At each time $k$, a message containing a measurement of the source, represented by $\check{x}_{c,k}$, can be sent over the broadcast channel to the monitors, where state estimates, represented by $\hat{x}_{i,k}$ for $i \in \mathcal{M}$, should be computed in a causal manner and over a finite time horizon~$N$.

The broadcast channel is modeled as a time-varying packet-erasure channel with packet error rate $\lambda_{i,k}$ for the $i$th link connecting the source to the $i$th monitor, and satisfies the input-output relation
\begin{align}\label{eq:broadcast}
z_{i,k+1} = \left\{
  \begin{array}{l l}
     \check{x}_{c,k}, & \ \text{if} \ u_{c,k} = 1 \ \wedge \ \gamma_{i,k} =1, \\[1\jot]
     \mathfrak{E}, & \ \text{otherwise}
  \end{array} \right.
\end{align}
for $k \in \mathbb{N}_{[0,N]}$ and $i \in \mathcal{M}$ with $z_{i,0} = \mathfrak{E}$ by convention, where $z_{i,k}$ is the output of the $i$th link, $u_{c,k} \in \{0,1\}$ is a binary variable such that $u_{c,k} = 1$ if a message containing $\check{x}_{c,k}$ is transmitted by the encoder at time $k$, and $u_{c,k} = 0$ otherwise; $\gamma_{i,k} \in \{0,1\}$ is a binary random variable such that $\gamma_{i,k} = 0$ if a packet loss occurs in the $i$th link at time $k$, and $\gamma_{i,k} = 1$ otherwise; and $\mathfrak{E}$ is a symbol representing packet loss or absence of transmission. It is assumed that the packet error rates $\lambda_{i,k}$ for $k \in \mathbb{N}_{[0,N]}$ are random variables forming a Markov chain; the packet error rate $\lambda_{i,k}$ is known at the encoder at each time $k$; the random variables $\gamma_{i,k}$ for $k \in \mathbb{N}_{[0,N]}$ are mutually independent given the respective packet error rates; measurement quantization error is negligible; and packet acknowledgments are sent back from the decoders to the encoder via ideal feedback links. 

The source is modeled as a partially observable Gauss--Markov process, satisfying the state and output equations
\begin{align}
	x_{c,k+1} &= A_{c,k} x_{c,k} + w_{c,k}\label{eq:sysA}\\[1.5\jot]
	y_{c,k} &= C_{c,k} x_{c,k} + v_{c,k} \label{eq:sensA}
\end{align}
for $k \in \mathbb{N}_{[0,N]}$ with initial condition $x_{c,0}$, where $x_{c,k} \in \mathbb{R}^n$ is the state of the source; $A_{c,k} \in \mathbb{R}^{n \times n}$ is the state matrix; $w_{c,k} \in \mathbb{R}^n$ is a Gaussian white noise with zero mean and covariance $W_{c,k} \succ 0$, $y_{c,k} \in \mathbb{R}^m$ is the output of the source; $C_{c,k} \in \mathbb{R}^{m \times n}$ is the output matrix, and $v_{c,k} \in \mathbb{R}^m$ is a Gaussian white noise with zero mean and covariance $V_{c,k} \succ 0$. It is assumed that the initial condition $x_{c,0}$ is a Gaussian vector with mean $m_{c,0}$ and covariance $M_{c,0}$; and the random variables $x_{c,0}$, $w_{c,t}$, and $v_{c,s}$ for $t,s \in \mathbb{N}_{[0,N]}$ are mutually independent.

Let $o_{i,k} := (y_{c,k}, z_{i,k}, \lambda_{i,k}, u_{c,k-1}, \gamma_{i,k-1})$ for $k \in \mathbb{N}_{[0,N]}$ and $i \in \mathcal{M}$. The information set of the common encoder at time $k$ can be represented by $\mathcal{I}^{e_c}_k = \{ o_{i,t} | t \in \mathbb{N}_{[0,k]},  i \in \mathcal{M} \}$ and that of the $i$th decoder by $\mathcal{I}^{d_i}_k = \{ z_{i,t} | t \in \mathbb{N}_{[0,k]} \}$, for $k \in \mathbb{N}_{[0,N]}$ and $i \in \mathcal{M}$. At each time $k$ for $k \in \mathbb{N}_{[0,N]}$, the common encoder must decide about $u_{c,k}$ and the $i$th decoder about $\hat{x}_{i,k}$ based on the Borel measurable stochastic kernels $\ProbM(u_{c,k} | \mathcal{I}^{e_c}_k)$ and $\ProbM(\hat{x}_{i,k} | \mathcal{I}^{d_i}_k)$, respectively. A coding policy profile $(\epsilon,\delta)$ consisting of an encoding policy $\epsilon$ and a decoding policy~$\delta$ is considered admissible if $\epsilon = \{ \ProbM(u_{c,k} | \mathcal{I}^{e_c}_k) | k \in \mathbb{N}_{[0,N]} \}$ and $\delta = \{ \ProbM(\hat{x}_{i,k} | \mathcal{I}^{d_i}_k) | k \in \mathbb{N}_{[0,N]}, i\in\mathcal{M} \}$. The first problem that we will address is to identify the best possible solution $(\epsilon^\star,\delta^\star)$ to the stochastic optimization problem
\begin{flalign}\label{eq:main_problemA}
\qquad \textit{Problem 1:} \qquad \qquad \underset{\epsilon \in \mathcal{E}, \delta \in \mathcal{D}}{\minimize} \ \Phi(\epsilon,\delta) &&
\end{flalign}
subject to the channel model in (\ref{eq:broadcast}), and the source model in (\ref{eq:sysA}) and (\ref{eq:sensA}), where $\mathcal{E}$ and $\mathcal{D}$ are the sets of admissible encoding policies and admissible decoding policies, respectively, and 
\begin{align}\label{eq:loss-functionA}
\Phi(\epsilon,\delta) := \E \bigg[ \sum_{k=0}^{N} \theta_{c,k} u_{c,k} \! +\! \sum_{k=0}^{N} \sum_{i\in\mathcal{M}} \omega_{i,k} e_{i,k}^T e_{i,k} \bigg]
\end{align}
for the estimation error $e_{i,k} := x_{c,k} - \hat{x}_{i,k}$, the weighting coefficient $\theta_{c,k} \geq 0$, which represents the cost of using the broadcast channel at time $k$, and the weighting coefficient $\omega_{i,k} \geq 0$, which specifies the importance of the task associated with the $i$th monitor at time $k$.

\subsection{State Estimation over a Multi-Access Channel}
Consider a networked system composed of two distinct stochastic sources with encoders sharing information among themselves, a common remote monitor with dedicated decoders corresponding to the sources, and a multi-access channel that connects the sources to the monitor. Let $\mathcal{M}$ be the set of sources. At each time $k$, a message containing a measurement of one of the sources, represented by $\check{x}_{j,k}$ for $j \in \mathcal{M}$, and its index, $j$, can be sent over the multi-access channel to the monitor, where state estimates, represented by $\hat{x}_{j,k}$ for $j \in \mathcal{M}$, should be computed in a causal way and over a finite time horizon~$N$.

The multi-access channel is modeled as a time-varying packet-erasure channel with packet error rate $\lambda_{j,k}$ for the $j$th link connecting the $j$th source to the monitor, satisfying the input-output relation
\begin{align}\label{eq:multiaccess}
z_{j,k+1} = \left\{
  \begin{array}{l l}
     (\check{x}_{j,k}, j), & \ \text{if} \ u_{j,k} = 1 \ \wedge \ \gamma_{j,k} =1, \\[1\jot]
     \mathfrak{E}, & \ \text{otherwise}
  \end{array} \right.
\end{align}
for $k \in \mathbb{N}_{[0,N]}$ and $j \in \mathcal{M}$ with $z_0 = \mathfrak{E}$ by convention, where $z_{j,k}$ is the output of the $j$th link, $u_{j,k} \in \{0,1\}$ is a binary variable such that $u_{j,k} = 1$ if a message containing $(\check{x}_{j,k}, j)$ is transmitted by the $j$th encoder at time $k$, and $u_{j,k} = 0$ otherwise; $\gamma_{j,k} \in \{0,1\}$ is a binary random variable such that $\gamma_{j,k} = 0$ if a packet loss occurs in the $j$th link at time $k$, and $\gamma_{j,k} = 1$ otherwise; and $\mathfrak{E}$ is a symbol representing packet loss or absence of transmission. It is assumed that the packet error rates $\lambda_{j,k}$ for $k \in \mathbb{N}_{[0,N]}$ are random variables forming a Markov chain; the packet error rate $\lambda_{j,k}$ is known at the encoders at each time $k$; the random variables $\gamma_{j,k}$ for $k \in \mathbb{N}_{[0,N]}$ are mutually independent given the respective packet error rates; measurement quantization error is negligible; and packet acknowledgments are sent back from the decoders to the encoders via ideal feedback links.

The sources are modeled as partially observable Gauss--Markov processes, satisfying the state and output equations
\begin{align}
	x_{j,k+1} &= A_{j,k} x_{j,k} + w_{j,k}\label{eq:sysB}\\[1.5\jot]
	y_{j,k} &= C_{j,k} x_{j,k} + v_{j,k} \label{eq:sensB}
\end{align}
for $k \in \mathbb{N}_{[0,N]}$ and $j \in \mathcal{M}$ with initial condition $x_{j,0}$, where $x_{j,k} \in \mathbb{R}^n$ is the state of the source; $A_{j,k} \in \mathbb{R}^{n \times n}$ is the state matrix; $w_{j,k} \in \mathbb{R}^n$ is a Gaussian white noise with zero mean and covariance $W_{j,k} \succ 0$; $y_{j,k} \in \mathbb{R}^m$ is the output of the source; $C_{j,k} \in \mathbb{R}^{m \times n}$ is the output matrix; and $v_{j,k} \in \mathbb{R}^m$ is a Gaussian white noise with zero mean and covariance $V_{j,k} \succ 0$. It is assumed that $x_{j,0}$ is a Gaussian vector with mean $m_{j,0}$ and covariance $M_{j,0}$; and $x_{j,0}$, $w_{j,t}$, and $v_{j,s}$ are mutually independent for all $t, s \in \mathbb{N}_{[0,N]}$.

Let $o_{j,k} := (y_{j,k}, z_{j,k}, \lambda_{j,k}, u_{j,k-1}, \gamma_{j,k-1})$ for $k \in \mathbb{N}_{[0,N]}$ and $j \in \mathcal{M}$. The information set of the $j$th encoder at time $k$ can be represented by $\mathcal{I}^{e_j}_k = \{ o_{j,t} | t \in \mathbb{N}_{[0,k]} \} \cup \mathcal{I}^{e_{\bar{j}}}_k$ and that of the $j$th decoder by $\mathcal{I}^{d_j}_k =  \{ z_{j,t} | t \in \mathbb{N}_{[0,k]} \}$, for $k \in \mathbb{N}_{[0,N]}$ and $j \in \mathcal{M}$. Note that this information structure is equivalent to that when there exists a network coordinator to which the sources report their local information. At each time $k$ for $k \in \mathbb{N}_{[0,N]}$, the $j$th encoder must decide about $u_{j,k}$ subject to the constraint $\sum_{j\in\mathcal{M}} u_{j,k} \leq 1$ and the $j$th decoder about $\hat{x}_{j,k}$ based on the Borel measurable stochastic kernels $\ProbM(u_{j,k} | \mathcal{I}^{e_j}_k)$ and $\ProbM(\hat{x}_{j,k} | \mathcal{I}^{d_j}_k)$, respectively. A coding policy profile $(\epsilon,\delta)$ consisting of an encoding policy $\epsilon$ and a decoding policy $\delta$ is considered admissible if $\epsilon = \{ \ProbM(u_{j,k} | \mathcal{I}^{e_j}_k) | \sum_{j\in\mathcal{M}} u_{j,k} \leq 1, k \in \mathbb{N}_{[0,N]}, j\in\mathcal{M} \}$ and $\delta = \{ \ProbM(\hat{x}_{j,k} | \mathcal{I}^{d_j}_k) | k \in \mathbb{N}_{[0,N]}, j\in\mathcal{M} \}$. The second problem that we will address is to identify the best possible solution $(\epsilon^\star,\delta^\star)$ to the stochastic optimization problem
\begin{flalign}\label{eq:main_problemB}
\qquad \textit{Problem 2:} \qquad \qquad \underset{\epsilon \in \mathcal{E}, \delta \in \mathcal{D}}{\minimize} \ \Phi(\epsilon,\delta) &&
\end{flalign}
subject to the channel model in (\ref{eq:multiaccess}), and the source model in (\ref{eq:sysB}) and (\ref{eq:sensB}), where $\mathcal{E}$ and $\mathcal{D}$ are the sets of admissible encoding policies and admissible decoding policies, respectively, and 
\begin{align}\label{eq:loss-functionB}
\Phi(\epsilon,\delta) := \E \bigg[ \sum_{k=0}^{N} \sum_{j\in\mathcal{M}} \theta_{j,k} u_{j,k} \! + \! \sum_{k=0}^{N}\sum_{j\in\mathcal{M}} \omega_{j,k} e_{j,k}^T e_{j,k} \bigg]
\end{align}
for the estimation error $e_{j,k} := x_{j,k} - \hat{x}_{j,k}$, the weighting coefficient $\theta_{j,k} \geq 0$, which represents the cost of using the multi-access channel at time $k$, and the weighting coefficient $\omega_{j,k} \geq 0$, which specifies the importance of the task associated with the $j$th source at time $k$.

\section{Main Results}\label{sec:main-results}
In this section, we present our main theoretical results. It is important to acknowledge that Problems 1 and 2 are team decision-making problems with non-classical information structures subject to signaling effects. The information structure is non-classical because any decision of an encoder can change the information set of a decoder while the latter does not have access to the information used by the former to make that decision. Moreover, a signaling effect exists because implicit information can be exchanged between an encoder and a decoder even when no sensory information is successfully communicated. It has been recognized in the literature that decision-making problems with such properties are significantly challenging~\cite{tsitsiklis1985, wu2013}.

The following definitions capture the notions of global optimality and value functions associated with Problems~1~and~2, and that of an access function, an operator for refining the valuation of information in multi-terminal settings.

\begin{definition}[Global optimality]
A policy profile $(\epsilon^\star,\delta^\star)$ in Problem 1 or 2 is globally optimal if
\begin{align*}
	\Phi(\epsilon^\star,\delta^\star) \leq \Phi(\epsilon,\delta), \ \text{for all } \epsilon \in \mathcal{E}, \delta \in \mathcal{D}.
\end{align*}
\end{definition}
\vspace{1mm}

\begin{definition}[Value function]
The value functions $V^c_k(\mathcal{I}^{e_c}_k)$ and $V^j_k(\mathcal{I}^{e_j}_k)$ associated with the loss function $\Phi(\epsilon,\delta)$ in Problems 1 and 2, respectively, are defined by
\begin{align}
	V^c_k(\mathcal{I}^{e_c}_k) :=& \min_{\epsilon \in \mathcal{E} : \delta = \delta^\star}\E \bigg[ \sum_{t=k}^{N-1} \theta_{c,t} u_{c,t} \nonumber\\
	&\qquad + \sum_{t=k}^{N-1} \sum_{i\in\mathcal{M}} \omega_{i,t+1} e_{i,t+1}^T e_{i,t+1} \Big| \mathcal{I}^{e_c}_k \bigg] \label{eq:Ve-def-bc}
\end{align}
and
\begin{align}
	V^j_k(\mathcal{I}^{e_j}_k) :=& \min_{\epsilon \in \mathcal{E} : \delta = \delta^\star}\E \bigg[ \sum_{t=k}^{N-1} \sum_{j\in\mathcal{M}} \theta_{j,t} u_{j,t} \nonumber\\
	&\qquad + \sum_{t=k}^{N-1} \sum_{j\in\mathcal{M}} \omega_{j,t+1} e_{j,t+1}^T e_{j,t+1} \Big| \mathcal{I}^{e_j}_k \bigg] \label{eq:Ve-def-mac}
\end{align}
for $k \in \mathbb{N}_{[0,N]}$.
\end{definition}

\begin{definition}[Access function]
The access function $\phi_y(x): \mathbb{R} \to \mathbb{R}$ is a scalar function defined as
\begin{align}
	\phi_y(x) = \left\{
  \begin{array}{l l}
     x, & \ \text{if} \ y > 0, \\[1\jot]
     -\infty, & \ \text{otherwise} .
  \end{array} \right.
\end{align}
\end{definition}
\vspace{1mm}

Associated with Problem 1, let us define the innovation at the common encoder $\nu_{c,k} := y_{c,k} - C_{c,k} \E [x_{c,k} | \mathcal{I}^{e_c}_{k-1}]$, the estimation error at the $i$th decoder based on the conditional mean $\hat{e}_{i,k} := x_{c,k} - \E[x_{c,k} | \mathcal{I}^{d_i}_k]$, the estimation mismatch at the $i$th decoder based on the conditional means $\tilde{e}_{i,k} := \E[x_{c,k} | \mathcal{I}^{e_c}_k] - \E[x_{c,k} | \mathcal{I}^{d_i}_k]$, the packet success rate in the $i$th link $\lambda'_{i,k} := 1- \lambda_{i,k}$, and the value difference $\Delta_{c,k} := \E[ V^c_{k+1}(\mathcal{I}^{e_c}_{k+1}) | \mathcal{I}^{e_c}_k, u_{c,k} = 0] - \E[ V^c_{k+1}(\mathcal{I}^{e_c}_{k+1}) | \mathcal{I}^{e_c}_k, u_{c,k} = 1]$. Furthermore, associated with Problem 2, let us define the innovation at the $j$th encoder $\nu_{j,k} := y_{j,k} - C_{j,k} \E [x_{j,k} | \mathcal{I}^{e_j}_{k-1}]$, the estimation error at the $j$th decoder based on the conditional mean $\hat{e}_{j,k} := x_{j,k} - \E[x_{j,k} | \mathcal{I}^{d_j}_k]$, the estimation mismatch at the $j$th decoder based on the conditional means $\tilde{e}_{j,k} := \E[x_{j,k} | \mathcal{I}^{e_j}_k] - \E[x_{j,k} | \mathcal{I}^{d_j}_k]$, the packet success rate in the $j$th link $\lambda'_{j,k} := 1- \lambda_{j,k}$, the value difference type one $\Delta^I_{j,k} := \E[ V^j_{k+1}(\mathcal{I}^{e_j}_{k+1}) | \mathcal{I}^{e_j}_k, u_{j,k} = 0, u_{\bar{j},k} = 0] - \E[ V^j_{k+1}(\mathcal{I}^{e_j}_{k+1}) | \mathcal{I}^{e_j}_k, u_{j,k} = 1, u_{\bar{j},k} = 0]$ for $j \in \mathcal{M}$, and the value difference type two $\Delta^{II}_{j,k} := \E[ V^j_{k+1}(\mathcal{I}^{e_j}_{k+1}) | \mathcal{I}^{e_j}_k, u_{j,k} = 0, u_{\bar{j},k} = 1] - \E[ V^j_{k+1}(\mathcal{I}^{e_j}_{k+1}) | \mathcal{I}^{e_j}_k, u_{j,k} = 1, u_{\bar{j},k} = 0]$ for $j \in \mathcal{M}$.

The most significant results of this article are given by the next theorems, which provide globally optimal solutions to Problems 1 and 2.

\begin{theorem}\label{thm:1}\emph{
For Problem~1, a globally optimal encoding policy $\epsilon^\star$ is specified by
\begin{align}
	u_{c,k} = \mathds{1}_{\phi_1(\chi_{c,k} - \theta_{c,k}) \geq 0}
\end{align}
in conjunction with $\check{x}_{c,k} = \E[x_{c,k} | \mathcal{I}^{e_c}_k]$  for $k \in \mathbb{N}_{[1,N]}$, where $\chi_{c,k} = \sum_{i \in \mathcal{M}} \lambda'_{i,k} \omega_{i,k+1} \tilde{e}_{i,k}^T A_{c,k}^T A_{c,k} \tilde{e}_{i,k} + \Delta_{c,k}$ is a symmetric function of $\tilde{e}_{i,k}$, which requires solving
\begin{align}
	\check{x}_{c,k} &= A_{c,k-1} \check{x}_{c,k-1} + K_{c,k} \nu_{c,k} \label{eq:est-KF-xhat} \\[1.5\jot]
	Q_{c,k} &= \big( (A_{c,k-1} Q_{c,k-1} A_{c,k-1}^T  \nonumber \\[1.5\jot]
	&\qquad \qquad \qquad + W_{c,k-1})^{-1} + C_{c,k}^T V_{c,k}^{-1} C_{c,k} \big)^{-1}\\[1.5\jot]
	\tilde{e}_{i,k} &= (1- u_{c,k-1} \gamma_{i,k-1}) A_{c,k-1} \tilde{e}_{i,k-1} + K_{c,k} \nu_{c,k} \label{eq:typeC:et}
\end{align}
for $k \in \mathbb{N}_{[1,N]}$ and $i \in \mathcal{M}$ with initial conditions $\check{x}_{c,0} = m_{c,0} + K_{c,0} \nu_{c,0}$, $Q_{c,0} = (M_{c,0}^{-1} + C_{c,0}^T V_{c,0}^{-1} C_{c,0})^{-1}$, and $\tilde{e}_{i,0} = K_{c,0} \nu_{c,0}$, where $K_{c,k} = Q_{c,k} C_{c,k}^T V_{c,k}^{-1}$; and a globally optimal decoding policy $\delta^\star$ is specified by
\begin{align}
	\hat{x}_{i,k} &= A_{c,k-1} \hat{x}_{i,k-1} \nonumber\\[1.6\jot]
	&\qquad + u_{c,k-1} \gamma_{i,k-1} A_{c,k-1} (\check{x}_{c,k-1} - \hat{x}_{i,k-1}) \label{eq:typeC:xh}
\end{align}
for $k \in \mathbb{N}_{[1,N]}$ and $i \in \mathcal{M}$ with initial condition $\hat{x}_{i,0} = m_{i,0}$, where $\hat{x}_{i,k} = \E[x_{c,k} | \mathcal{I}^{d_i}_k]$.}
\end{theorem}
\begin{IEEEproof}
See Section~\ref{sec:der-main-results}.
\end{IEEEproof}

\begin{theorem}\label{thm:2}\emph{
For Problem~2, a globally optimal encoding policy $\epsilon^\star$ is specified by
\begin{align}
	u_{j,k} = \mathds{1}_{\phi_{\rho_{j,k}}(\chi_{j,k} - \theta_{j,k}) \geq 0}
\end{align}
in conjunction with $\check{x}_{j,k} = \E[x_{j,k} | \mathcal{I}^{e_j}_k]$ for $k \in \mathbb{N}_{[1,N]}$ and $j \in \mathcal{M}$, where $\chi_{j,k} = \lambda'_{j,k} \omega_{j,k+1} \tilde{e}_{j,k}^T A_{j,k}^T A_{j,k} \tilde{e}_{j,k} + \Delta^I_{j,k}$ and $\rho_{j,k} = \lambda'_{j,k} \omega_{j,k+1} \tilde{e}_{j,k}^T A_{j,k}^T A_{j,k} \tilde{e}_{j,k} - \lambda'_{\bar{j},k} \omega_{\bar{j},k+1} \tilde{e}_{\bar{j},k}^T A_{\bar{j},k}^T A_{\bar{j},k} \tilde{e}_{\bar{j},k} + \Delta^{II}_{j,k}$ are symmetric functions of $\tilde{e}_{j,k}$, which requires solving
\begin{align}
	\check{x}_{j,k} &= A_{j,k-1} \check{x}_{j,k-1} + K_{j,k} \nu_{j,k} \label{eq:est-KF-xhat} \\[1.5\jot]
	Q_{j,k} &= \big( (A_{j,k-1} Q_{j,k-1} A_{j,k-1}^T  \nonumber\\[1.5\jot]
	&\qquad \qquad \qquad + W_{j,k-1})^{-1} + C_{j,k}^T V_{j,k}^{-1} C_{j,k} \big)^{-1}\\[1.5\jot]
	\tilde{e}_{j,k} &= (1- u_{j,k-1} \gamma_{j,k-1}) A_{j,k-1} \tilde{e}_{j,k-1} + K_{j,k} \nu_{j,k} \label{eq:typeC:et}
\end{align}
for $k \in \mathbb{N}_{[1,N]}$ and $j \in \mathcal{M}$ with initial conditions $\check{x}_{j,0} = m_{j,0} + K_{j,0} \nu_{j,0}$, $Q_{j,0} = (M_{j,0}^{-1} + C_{j,0}^T V_{j,0}^{-1} C_{j,0})^{-1}$, and $\tilde{e}_{j,0} = K_{j,0} \nu_{j,0}$, where $K_{j,k} = Q_{j,k} C_{j,k}^T V_{j,k}^{-1}$; and a globally optimal decoding policy $\delta^\star$ is specified by
\begin{align}
	\hat{x}_{j,k} &= A_{j,k-1} \hat{x}_{j,k-1} \nonumber\\[1.6\jot]
	&\qquad + u_{j,k-1} \gamma_{j,k-1} A_{j,k-1} (\check{x}_{j,k-1} - \hat{x}_{j,k-1}) \label{eq:typeC:xh}
\end{align}
for $k \in \mathbb{N}_{[1,N]}$ and $j \in \mathcal{M}$ with initial condition $\hat{x}_{j,0} = m_{j,0}$, where $\hat{x}_{j,k} = \E[x_{j,k} | \mathcal{I}^{d_j}_k]$.}
\end{theorem}
\begin{IEEEproof}
See Section~\ref{sec:der-main-results}.
\end{IEEEproof}

\begin{remark}
The results introduce two novel semantic metrics. The first metric arising in Problem 1 is the ``\emph{dissemination value of information}'', i.e., $\phi_1(\chi_{c,k} - \theta_{c,k})$, which quantifies the valuation of provisioning a piece of information to multiple receivers simultaneously. This quantity is obtained when the access function is the identity function. Accordingly, a message at time $k$ is transmitted over the broadcast channel only if it is \emph{valuable in aggregate} for multiple receivers, i.e., only if the dissemination value of information at time $k$ is nonnegative. The second metric arising in Problem 2 is the ``\emph{prioritization value of information}'', i.e., $\phi_{\rho_{j,k}}(\chi_{j,k} - \theta_{j,k})$, which quantifies the valuation of provisioning a piece of information chosen from one out of multiple transmitters. This quantity is obtained when the access function is defined with respect to $\rho_{j,k}$. Accordingly, a message at time $k$ is transmitted over the multi-access channel only if it is \emph{both urgent and valuable}, i.e., only if the prioritization value of information at time $k$ is nonnegative.
\end{remark}

\begin{remark}
Note that, in practice, the terms $\chi_{c,k}$ for Problem 1 and the terms $\chi_{j,k}$ and $\rho_{j,k}$ for Problem 2 can be approximated based on the one-step lookahead algorithm (see, e.g., \cite{bertsekas1995DP}). Using this procedure, we obtain $\chi_{c,k} \simeq \sum_{i \in \mathcal{M}} \lambda'_{i,k} \omega_{i,k+1} \tilde{e}_{i,k}^T A_{c,k}^T A_{c,k} \tilde{e}_{i,k}$ for Problem 1, and $\chi_{j,k} \simeq \lambda'_{j,k} \omega_{j,k+1} \tilde{e}_{j,k}^T A_{j,k}^T A_{j,k} \tilde{e}_{j,k}$ and $\rho_{j,k} \simeq \lambda'_{j,k} \omega_{j,k+1} \tilde{e}_{j,k}^T A_{j,k}^T A_{j,k} \tilde{e}_{j,k} - \lambda'_{\bar{j},k} \omega_{\bar{j},k+1} \tilde{e}_{\bar{j},k}^T A_{\bar{j},k}^T A_{\bar{j},k} \tilde{e}_{\bar{j},k}$ for Problem 2. As mentioned, our results are for $|\mathcal{M}| = 2$. For the general case $|\mathcal{M}| > 2$, it is anticipated that a few modifications in the structures of the characterized optimal policies will be required. In particular, the term $\Delta_{c,k}$ for Problem 1, and the terms $\Delta^I_{j,k}$ and $\Delta^{II}_{j,k}$ and the function $\phi_y(x)$ for Problem 2 should be amended in a way that the effects of all sources or monitors are captured. 
\end{remark}

\begin{remark}
Note that an admissible encoding policy is in general defined based on a condition like $f_k(\mathcal{I}^{e_s}_k) \in \mathcal{F}_k$ for an appropriate index $s$, where $f_k(.)$ and $\mathcal{F}_k$ are a measurable function and a measurable set, respectively (see, e.g., \cite{sijs2012}). As a result, characterizing such a policy typically necessitates complex computations. Our results, however, indicate that there exists a globally optimal encoding policy $\epsilon^\star$ that is of a threshold type. This structure significantly simplifies the design of the encoding policy. Furthermore, note that an admissible decoding policy is in general dependent on the conditional distribution $\ProbM(x_k | \mathcal{I}^{d_s}_k)$ for an appropriate index $s$, which is non-Gaussian due to the signaling effect (see, e.g., \cite{wu2013}). As a result, such a policy is typically nonlinear and without any analytical form. Our results, however, indicate that there exists a globally optimal decoding policy $\delta^\star$ that is linear without being influenced by the signaling effect. This structure dramatically simplifies the design of the decoding~policy.
\end{remark}

The next corollaries present the results of Theorems~\ref{thm:1}~and~\ref{thm:2} for the special case of one-shot communication, i.e., when $N = 1$. In this case, one only needs to determine $u_{c,0}$ and $\hat{x}_{i,1}$ for $i \in \mathcal{M}$ in Problem 1, and $u_{j,1}$ and $\hat{x}_{j,1}$ for $j \in \mathcal{M}$ in Problem 2.

\begin{corollary}\label{coro:1}\emph{
For Problem~1 with the time horizon $N=1$, a globally optimal encoding policy $\epsilon^\star$ is specified by
\begin{align}
	u_{c,0} = \mathds{1}_{\phi_1(\chi_{c,0} - \theta_{c,0}) \geq 0}
\end{align}
in conjuction with $\check{x}_{c,0} = m_{c,0} + K_{c,0} \nu_{c,0}$, where $\chi_{c,0} = (\lambda'_{1,0} \omega_{1,1} + \lambda'_{2,0}\omega_{2,1}) \nu_{c,0}^T K_{c,0}^T A_{c,0}^T A_{c,0} K_{c,0} \nu_{c,0}$, and a globally optimal decoding policy $\delta^\star$ is specified by
\begin{align}
	\hat{x}_{i,1} = A_{c,0} m_{i,0} + u_{c,0} \gamma_{i,0} A_{c,0} K_{c,0} \nu_{c,0} \label{eq:typeC:xh}
\end{align}
for $i \in \mathcal{M}$, where $\hat{x}_{i,1} = \E[x_{c,1} | \mathcal{I}^{d_i}_k]$ and $\nu_{c,0} = y_{c,0} - C_{c,0} m_{i,0}$.}
\end{corollary}
\begin{IEEEproof}
See Section~\ref{sec:der-main-results}.
\end{IEEEproof}

\begin{corollary}\label{coro:2}\emph{
For Problem~2 with the time horizon $N=1$, a globally optimal encoding policy $\epsilon^\star$ is specified~by
\begin{align}
	u_{j,0} = \mathds{1}_{\phi_{\rho_{j,0}} (\chi_{j,0} - \theta_{j,0}) \geq 0}
\end{align}
in conjuction with $\check{x}_{j,0} = m_{j,0} + K_{j,0} \nu_{j,0}$ for $j \in \mathcal{M}$, where $\chi_{j,0} = \lambda'_{j,0} \omega_{j,1} \nu_{j,0}^T K_{j,0}^T A_{j,0}^T A_{j,0} K_{j,0} \nu_{j,0}$ and $\rho_{j,0} = \chi_{j,0} - \chi_{\bar{j},0}$, and a globally optimal decoding policy $\delta^\star$ is specified~by
\begin{align}
	\hat{x}_{j,1} = A_{j,0} m_{j,0} + u_{j,0} \gamma_{j,0} A_{j,0} K_{j,0} \nu_{j,0} \label{eq:typeC:xh}
\end{align}
for $j \in \mathcal{M}$, where $\hat{x}_{j,1} = \E[x_{j,1} | \mathcal{I}^{d_j}_k]$ and $\nu_{j,0} = y_{j,0} - C_{j,0} m_{j,0}$.}
\end{corollary}
\begin{IEEEproof}
See Section~\ref{sec:der-main-results}.
\end{IEEEproof}

\begin{remark}
The results for the time horizon $N=1$ are of interest as they analytically illustrate the globally optimal solutions and enhance our understanding of their structures. Note that, in this case, the encoding and decoding policies are expressed in terms of $\nu_{c,0}$ and $\lambda_{1,0}$ and $\lambda_{2,0}$ for Problem~1, and of $\nu_{1,0}$, $\nu_{2,0}$, $\lambda_{1,0}$, and $\lambda_{2,0}$ for Problem~2. Moreover, note that, for any fixed $\nu_{c,0}$, $\nu_{1,0}$, $\nu_{2,0}$, and $\theta_{c,0}$, $\theta_{1,0}$, and $\theta_{2,0}$, there exist cutoff values for $\lambda_{1,0}$ and $\lambda_{2,0}$ below which $u_{c,0}$, $u_{1,0}$, or $u_{2,0}$ becomes zero. This implies that when the channel conditions are poor no message should be transmitted over the channel. The adaptiveness of the optimal encoding policies to the channel conditions here in fact resembles that of the optimal transmit power policy to the channel condition in \cite{goldsmith1997-1, goldsmith1997-2}, where it was shown that when the channel condition is below a cutoff value no data should be transmitted over the channel. Nevertheless, the focus of the above-mentioned studies is on a tradeoff between the average transmit power and the fading channel capacity.
\end{remark}

\section{Derivation of Main Results}\label{sec:der-main-results}
This section is dedicated to the derivation of the main results. First, we provide the proof of Theorems~\ref{thm:1} and \ref{thm:2}.

\begin{IEEEproof}
The proof is organized in four steps.

\underline{Step 0.}
For Problem 1, assuming a globally optimal encoding policy is implemented, the optimal value that minimizes the MSE at time $k$ for the $i$th decoder, given $\mathcal{I}^{d_i}_k$, is $\E[x_{c,k} | \mathcal{I}^{d_i}_k]$. Additionally, $\E[x_{c,k} | \mathcal{I}^{e_c}_k]$ combines all current and previous outputs of the common source that are accessible to the common encoder at time $k$. If this fused measurement is transmitted successfully, the $i$th decoder can develop a state estimate that is equivalent to that when having access to all previous outputs of the source, resulting in the minimum possible MSE. Therefore, without loss of optimality, $\check{x}_{c,k} = \E[x_{c,k} | \mathcal{I}^{e_c}_k]$ can be adopted as the message transmitted by the common encoder, and $\hat{x}_{i,k} = \E[x_{c,k} | \mathcal{I}^{d_i}_k]$ as the state estimate computed by the $i$th decoder.

Similarly, for Problem 2, assuming a globally optimal encoding policy is implemented, the optimal value that minimizes the MSE at time $k$ for the $j$th decoder, given $\mathcal{I}^{d_j}_k$, is $\E[x_{j,k} | \mathcal{I}^{d_j}_k]$ for $j\in\mathcal{M}$. Additionally, $\E[x_{j,k} | \mathcal{I}^{e_j}_k]$ combines all current and previous outputs of the $j$th source that are accessible to the $j$th encoder at time $k$. If this fused measurement is transmitted successfully, the $j$th decoder can develop a state estimate that is equivalent to that when having access to all previous outputs of the $j$th source, resulting in the minimum possible MSE. Therefore, without loss of optimality, $\check{x}_{j,k} = \E[x_{j,k} | \mathcal{I}^{e_j}_k]$ can be adopted as the message transmitted by the $j$th encoder, and $\hat{x}_{j,k} = \E[x_{j,k} | \mathcal{I}^{d_j}_k]$ as the state estimates computed by the $j$th decoder.

\underline{Step 1.} For our analysis in what follows, we need to introduce hypothetically dummy entities. In particular, for Problem~1, instead of a common source and a common encoder, we consider two sources and two encoders that work exactly as the original source and the original encoder, respectively. This duplication will allow us to present our derivation in a way that is, unless otherwise stated, the same for both Problems 1 and 2. Accordingly, we will adopt the following conventions. For Problem 1, we introduce $\mathcal{I}^{e_c}_k = \mathcal{I}^{e_1}_k = \mathcal{I}^{e_2}_k$, $x_{c,k} = x_{1,k} = x_{2,k}$, $y_{c,k} = y_{1,k} = y_{2,k}$, $u_{c,k} = u_{1,k} = u_{2,k}$, $A_{c,k} = A_{1,k} = A_{2,k}$, $B_{c,k} = B_{1,k} = B_{2,k}$, $C_{c,k} = C_{1,k} = C_{2,k}$, $W_{c,k} = W_{1,k} = W_{2,k}$, $V_{c,k} = V_{1,k} = V_{2,k}$, $K_{c,k} = K_{1,k} = K_{2,k}$, $\nu_{c,k} = \nu_{1,k} = \nu_{2,k}$, $m_{c,k} = m_{1,k} = m_{2,k}$, and $M_{c,k} = M_{1,k} = M_{2,k}$.

We say a scheduling policy is innovation-based if, at each time $k$, it depends on $\boldsymbol{\nu}_{s,0:k}$ instead of $\mathbf{y}_{s,0:k}$ and $\mathbf{z}_{s,0:k}$. We show that $\Phi(\epsilon^\mathsf{n},\delta^\mathsf{o}) = \Phi(\epsilon^\mathsf{o},\delta^\mathsf{o})$, where $\epsilon^\mathsf{n}$ is an innovation-based scheduling policy. Let us define
\begin{align}
\mathcal{Z}_k &= \Big\{ o'_{s,t} \big| t \in \mathbb{N}_{[0,k]}, s \in \mathcal{M} \Big\},
\end{align}
where $o'_{s,t} := (z_{s,t}, \lambda_{s,t}, u_{s,t-1}, \gamma_{s,t-1})$. From the definition of the innovation and by Lemma~\ref{lem:estimator-encoder}, we can write $\mathbf{y}_{s,0:k} = \boldsymbol{\nu}_{s,0:k} + G_{s,k} \check{\mathbf{x}}_{s,0:k-1}$ and $\check{\mathbf{x}}_{s,0:k} = H_{s,k} \boldsymbol{\nu}_{s,0:k}$, where $G_{s,k}$ and $H_{s,k}$ are matrices of proper dimensions. Putting these equations together, we find $\mathbf{y}_{s,0:k} = \boldsymbol{\nu}_{s,0:k} + G_{s,k} H_{s,k-1} \boldsymbol{\nu}_{s,0:k-1}$. Therefore, $\Prob_{\epsilon^\mathsf{o}}(u_{s,k} | \mathbf{y}_{1,0:k}, \mathbf{y}_{2,0:k}, \mathcal{Z}_k)$ can equivalently be written as $\Prob_{\epsilon^\mathsf{n}}(u_{s,k} | \boldsymbol{\nu}_{1,0:k}, \boldsymbol{\nu}_{2,0:k}, \mathcal{Z}_k)$. This establishes that $\Phi(\epsilon^\mathsf{n},\delta^\mathsf{o}) = \Phi(\epsilon^\mathsf{o},\delta^\mathsf{o})$. As our subsequent analysis, for brevity, we write $\epsilon^\mathsf{n}$ as $\Prob_{\epsilon^\mathsf{n}}(u_{s,k} | \boldsymbol{\nu}_{s,0:k}, \mathbf{u}_{s,0:k-1}, \boldsymbol{\gamma}_{s,0:k-1})$, and omit the dependency of $\epsilon^\mathsf{n}$ on other variables.

\underline{Step 2.} Let $\mathcal{B}(r)$ be a ball of radius $r$ centered at the origin and of proper dimension. Define $\hbar_{s,k} := T_{s,k} \boldsymbol{\nu}_{s,0:k} \in \mathbb{R}^m$ for a given matrix $T_{s,k}$, and $h_{s,k} := u_{s,k} \gamma_{s,k}$. We show that $\Phi(\epsilon^\mathsf{s},\delta^\mathsf{o}) \leq \Phi(\epsilon^\mathsf{n},\delta^\mathsf{o})$, where $\epsilon^\mathsf{s}$ is a special form of $\epsilon^\mathsf{n}$ that is symmetric with respect to $\boldsymbol{\nu}_{s,0:k}$ for $s \in \mathcal{M}$ at each time $k$ and such that the following conditions are satisfied:
\begin{equation*}
\begin{aligned}
& \int_{\mathbb{R}^m} \big(\lambda_{s,k} + \lambda'_{s,k} \Prob_{\epsilon^\mathsf{s}}(u_{s,k} = 0| \hbar_{s,k}, \mathbf{h}_{s,0:k-1} = 0) \big)\\[0.5\jot]
&\qquad \qquad \qquad \qquad \qquad \quad \times \Prob_{\epsilon^\mathsf{s}}( \hbar_{s,k} | \mathbf{h}_{s,0:k-1} = 0) d\hbar_{s,k}\\[1\jot]
&= \int_{\mathbb{R}^m} \Big( \big(\lambda_{s,k} + \lambda'_{s,k} \Prob_{\epsilon^\mathsf{n}}(u_{s,k} = 0| \hbar_{s,k}, \mathbf{h}_{s,0:k-1} = 0) \big)\\
&\qquad \qquad \qquad \qquad \qquad \quad \times \Prob_{\epsilon^\mathsf{n}}(\hbar_{s,k} | \mathbf{h}_{s,0:k-1} = 0) \Big)^* d\hbar_{s,k}
\end{aligned}
\end{equation*}
and
\begin{equation*}
\begin{aligned}
& \int_{\mathcal{B}(r)} \big(\lambda_{s,k} + \lambda'_{s,k} \Prob_{\epsilon^\mathsf{s}}(u_{s,k} = 0| \hbar_{s,k}, \mathbf{h}_{s,0:k-1} = 0) \big)\\[0.5\jot]
&\qquad \qquad \qquad \qquad \qquad \quad \times \Prob_{\epsilon^\mathsf{s}}( \hbar_{s,k} | \mathbf{h}_{s,0:k-1} = 0) d\hbar_{s,k}\\[1\jot]
&= \int_{\mathcal{B}(r)} \Big( \big(\lambda_{s,k} + \lambda'_{s,k} \Prob_{\epsilon^\mathsf{n}}(u_{s,k} = 0| \hbar_{s,k}, \mathbf{h}_{s,0:k-1} = 0) \big)\\
&\qquad \qquad \qquad \qquad \qquad \quad \times \Prob_{\epsilon^\mathsf{n}}(\hbar_{s,k} | \mathbf{h}_{s,0:k-1} = 0) \Big)^* d\hbar_{s,k}
\end{aligned}
\end{equation*}
for all $r \geq 0$ with $( \lambda_{s,k} + \lambda'_{s,k} \Prob_{\epsilon^\mathsf{s}}(u_{s,k}= 0| \hbar_{s,k}, \mathbf{h}_{s,0:k-1} = 0) ) \Prob_{\epsilon^\mathsf{s}}( \hbar_{s,k} | \mathbf{h}_{s,0:k-1} = 0)$ as a radially symmetric function of $\hbar_{s,k}$. We know that $h_{s,k} = 0$ only when $(u_{s,k} = 1 \; \wedge \; \gamma_{s,k} = 0) \; \vee \; (u_{s,k} = 0)$. Accordingly, we can write
\begin{align*}
	&\Prob_{\epsilon^\mathsf{n}} (h_{s,k} = 0 \big| \hbar_{s,k}, \mathbf{h}_{s,0:k-1} = 0 ) \\[2.5\jot]
	&\  = \Prob_{\epsilon^\mathsf{n}} (u_{s,k} = 1 \big| \hbar_{s,k}, \mathbf{h}_{s,0:k-1} = 0 )\\[2.5\jot]
	&\qquad \times \Prob_{\epsilon^\mathsf{n}} (u_{s,k} \gamma_{s,k} = 0 \big| \hbar_{s,k}, \mathbf{h}_{s,0:k-1} = 0, u_{s,k} = 1 ) \\[2.5\jot]
	&\qquad + \Prob_{\epsilon^\mathsf{n}} (u_{s,k} = 0 \big| \hbar_{s,k}, \mathbf{h}_{s,0:k-1} = 0 )\\[2.5\jot]
	&\qquad \times  \Prob_{\epsilon^\mathsf{n}} (u_{s,k} \gamma_{s,k} = 0 \big| \hbar_{s,k}, \mathbf{h}_{s,0:k-1} = 0, u_{s,k} = 0 )\\[2.5\jot]
	&\ = \lambda_{s,k} \Prob_{\epsilon^\mathsf{n}} ( u_{s,k} = 1 \big| \hbar_{s,k}, \mathbf{h}_{s,0:k-1} = 0 ) \\[2.5\jot]
	&\qquad + \Prob_{\epsilon^\mathsf{n}} (u_{s,k} = 0 \big| \hbar_{s,k}, \mathbf{h}_{s,0:k-1} = 0 )\\[2.5\jot]
	&\ = \lambda_{s,k} + \lambda'_{s,k} \Prob_{\epsilon^\mathsf{n}} (u_{s,k} = 0 \big| \hbar_{s,k} , \mathbf{h}_{s,0:k-1} = 0 )
\end{align*}
where in the first equality we used the law of total probability and in the second equality the fact that $\gamma_{s,k}$ is independent of $\hbar_{s,k}$, $\mathbf{h}_{s,0:k-1}$, and $u_{s,k}$. 

Therefore, the above conditions can be written in the following equivalent form:
\begin{align}\label{eq:construction1B}
&\int_{\mathbb{R}^m} \Prob_{\epsilon^\mathsf{s}}(h_{s,k} = 0| \hbar_{s,k}, \mathbf{h}_{s,0:k-1} = 0) \nonumber\\ 
&\qquad \qquad \qquad \quad \times \Prob_{\epsilon^\mathsf{s}}( \hbar_{s,k} | \mathbf{h}_{s,0:k-1} = 0) d\hbar_{s,k} \nonumber\\[1.5\jot]
& = \int_{\mathbb{R}^m} \Big( \Prob_{\epsilon^\mathsf{n}}(h_{s,k} = 0| \hbar_{s,k}, \mathbf{h}_{s,0:k-1} = 0) \nonumber\\
&\qquad \qquad \qquad \quad \times \Prob_{\epsilon^\mathsf{n}}( \hbar_{s,k} | \mathbf{h}_{s,0:k-1} = 0) \Big)^* d\hbar_{s,k}
\end{align}
and
\begin{align}\label{eq:construction2B}
& \int_{\mathcal{B}(r)} \Prob_{\epsilon^\mathsf{s}}(h_{s,k} = 0| \hbar_{s,k}, \mathbf{h}_{s,0:k-1} = 0) \nonumber\\
&\qquad \qquad \qquad \quad \times \Prob_{\epsilon^\mathsf{s}}( \hbar_{s,k} | \mathbf{h}_{s,0:k-1} = 0) d\hbar_{s,k} \nonumber\\[1.5\jot]
& = \int_{\mathcal{B}(r)} \Big( \Prob_{\epsilon^\mathsf{n}}(h_{s,k} = 0| \hbar_{s,k}, \mathbf{h}_{s,0:k-1} = 0) \nonumber\\
&\qquad \qquad \qquad \quad \times \Prob_{\epsilon^\mathsf{n}}( \hbar_{s,k} | \mathbf{h}_{s,0:k-1} = 0) \Big)^*d\hbar_{s,k}
\end{align}
for all $r \geq 0$ with $\Prob_{\epsilon^\mathsf{s}}(h_{s,k}= 0| \hbar_{s,k}, \mathbf{h}_{s,0:k-1} = 0) \Prob_{\epsilon^\mathsf{s}}( \hbar_{s,k} | \mathbf{h}_{s,0:k-1} = 0)$ as a radially symmetric function of $\hbar_{s,k}$.

By Lemma~\ref{lem:equiv-loss}, to prove that $\Phi(\epsilon^\mathsf{s},\delta^\mathsf{o}) \leq \Phi(\epsilon^\mathsf{n},\delta^\mathsf{o})$, it suffices to prove that $\Omega^{s,M}_{\epsilon^\mathsf{s}}(\tilde{e}_{s,0}) \leq \Omega^{s,M}_{\epsilon^\mathsf{n}}(\tilde{e}_{s,0})$ for any $M \in \mathbb{N}_{[0,N]}$, any $s \in \mathcal{M}$, and any $\tilde{e}_{s,0}$. Observe that $\tilde{e}_{s,0} = K_{s,0} \nu_{s,0}$ is the same under both $\epsilon^\mathsf{n}$ and $\epsilon^\mathsf{s}$, and that $u_{s,0}$ has no effects on state estimation when the time horizon is zero. Hence, the claim holds for the time horizon zero. We assume that the claim also holds for all time horizons from $1$ to $M-1$, and will show that the terms in $\Omega^{s,M}_{\epsilon^\mathsf{n}}(\tilde{e}_{s,0})$ are not less than those in $\Omega^{s,M}_{\epsilon^\mathsf{s}}(\tilde{e}_{s,0})$ for the time horizon $M$.

First, for the probability coefficients, we have
\begin{align*}
&\Prob_{\epsilon^\mathsf{n}}(h_{s,k-1} = 0 | \mathbf{h}_{s,0:k-2} = 0)\\[1.25\jot]
&= \int_{\mathbb{R}^m} \Prob_{\epsilon^\mathsf{n}}(h_{s,k-1} = 0| \hbar_{s,k-1}, \mathbf{h}_{s,0:k-2} = 0)\\[1.25\jot]
&\qquad \qquad \qquad \times \Prob_{\epsilon^\mathsf{n}}(\hbar_{s,k-1} | \mathbf{h}_{s,0:k-2} = 0) d\hbar_{s,k-1}\\[1.25\jot]
&= \int_{\mathbb{R}^m} \Prob_{\epsilon^\mathsf{s}}(h_{s,k-1} = 0| \hbar_{s,k-1}, \mathbf{h}_{s,0:k-2} = 0)\\[1.25\jot]
&\qquad \qquad \qquad \times \Prob_{\epsilon^\mathsf{s}}(\hbar_{s,k-1} | \mathbf{h}_{s,0:k-2} = 0) d\hbar_{s,k-1}\\[2\jot]
&=\Prob_{\epsilon^\mathsf{s}}(h_{s,k-1} = 0 | \mathbf{h}_{s,0:k-2} = 0)
\end{align*}
where the second equality comes from (\ref{eq:construction1B}). This also implies that $\Prob_{\epsilon^\mathsf{n}}(\mathbf{h}_{s,0:k-1} = 0) = \Prob_{\epsilon^\mathsf{s}}(\mathbf{h}_{s,0:k-1} = 0) $ and that $\Prob_{\epsilon^\mathsf{n}}(\mathbf{h}_{s,0:k-1} = 0, h_{s,k} = 1) = \Prob_{\epsilon^\mathsf{s}}(\mathbf{h}_{s,0:k-1} = 0, h_{s,k} = 1)$. Moreover, for the terms involving the expected value of the transmission decision, we have
\begin{align*}
\E_{\epsilon^\mathsf{n}} \Big[ & u_{s,k} \big| \mathbf{h}_{s,0:k-1} = 0 \Big]\\[1\jot]
	&= 1 - \Prob_{\epsilon^\mathsf{n}}(u_{s,k} = 0| \mathbf{h}_{s,0:k-1} = 0)\\[1\jot]	
	&= \frac{1}{\lambda'_{s,k}} - \frac{1}{\lambda'_{s,k}} \Prob_{\epsilon^\mathsf{n}}(h_{s,k} = 0| \mathbf{h}_{s,0:k-1} = 0)\\[1\jot]
	&= \frac{1}{\lambda'_{s,k}} - \frac{1}{\lambda'_{s,k}} \Prob_{\epsilon^\mathsf{s}}(h_{s,k} = 0| \mathbf{h}_{s,0:k-1} = 0)
\end{align*}
\begin{align*}
	&= 1 - \Prob_{\epsilon^\mathsf{s}}(u_{s,k} = 0| \mathbf{h}_{s,0:k-1} = 0)\\[1\jot]
	&=\E_{\epsilon^\mathsf{s}} \Big[ u_{s,k} \big| \mathbf{h}_{s,0:k-1} = 0 \Big].
\end{align*}

Observe that, by Lemma~\ref{lem:mismatch-dyn-est}, when $\mathbf{h}_{s,0:k-1}=0$, $\tilde{e}_{s,t}$ satisfies
\begin{align*}
	\tilde{e}_{s,t} &= A_{s,t-1} \tilde{e}_{s,t-1} + K_{s,t} \nu_{s,t} - \imath_{s,t-1}
\end{align*}
for $t \in \mathbb{N}_{[1,k]}$ with initial condition $\tilde{e}_{s,0} = K_{s,0} \nu_{s,0}$. Accordingly, we can find a proper matrix $E_{s,k}$ and a proper vector $c_{s,k}$, both independent of $\boldsymbol{\nu}_{s,0:k}$, such that $\tilde{e}_{s,k} = E_{s,k} \boldsymbol{\nu}_{s,0:k-1} + K_{s,k} \nu_{s,k} + c_{s,k}$ under $\epsilon^\mathsf{n}$. We know that $\epsilon^\mathsf{s}$ is symmetric with respect to $\boldsymbol{\nu}_{s,0:k}$ at each time $k$. Therefore, by Lemma~\ref{lem:zero-residuals}, we deduce that $\tilde{e}_{s,k} = E_{s,k} \boldsymbol{\nu}_{s,0:k-1} + K_{s,k} \nu_{s,k}$ under $\epsilon^\mathsf{s}$. For the terms involving the expected value of the quadratic estimation mismatch, we can then write
\begin{align*}
	&\E_{\epsilon^\mathsf{n}} \Big[ \tilde{e}_{s,k}^T \tilde{e}_{s,k} \big| \mathbf{h}_{s,0:k-1} = 0 \Big]\\[1\jot]
	&\ = \E_{\epsilon^\mathsf{n}} \Big[ \big(E_{s,k} \boldsymbol{\nu}_{s,0:k-1} + K_{s,k} \nu_{s,k} + c_{s,k} \big)^T\\[1\jot]
	&\qquad \times \big(E_{s,k} \boldsymbol{\nu}_{s,0:k-1} + K_{s,k} \nu_{s,k} + c_{s,k} \big) \big| \mathbf{h}_{s,0:k-1} = 0 \Big]\\[1\jot]
	&\ = \E_{\epsilon^\mathsf{n}} \Big[ \big(E_{s,k} \boldsymbol{\nu}_{s,0:k-1} + c_{s,k} \big)^T \big(E_{s,k} \boldsymbol{\nu}_{s,0:k-1} + c_{s,k} \big)\\[1\jot]
	&\qquad + \nu_{s,k}^T K_{s,k}^T K_{s,k} \nu_{s,k} \big| \mathbf{h}_{s,0:k-1} = 0 \Big]
\end{align*}
where in the second equality we used the fact that $\nu_{s,k}$ has zero mean and is independent of $\boldsymbol{\nu}_{s,0:k-1}$ and $\mathbf{h}_{s,0:k-1}$. 

Choose $T_{s,k-1} = E_{s,k}$, and define $f_{\epsilon^\mathsf{n}}(\hbar_{s,k-1}, \nu_{s,k}) := (\hbar_{s,k-1} + c_{s,k})^T (\hbar_{s,k-1} + c_{s,k}) + \nu_{s,k}^T K_{s,k}^T K_{s,k} \nu_{s,k}$, $f_{\epsilon^\mathsf{s}}(\hbar_{s,k-1}, \nu_{s,k}) := \hbar_{s,k-1}^T \hbar_{s,k-1} + \nu_{s,k}^T K_{s,k}^T K_{s,k} \nu_{s,k}$, $g_{\epsilon^\mathsf{n}}(\hbar_{s,k-1}, \nu_{s,k}) := z -\min_z \{z,f_{\epsilon^\mathsf{n}}(\hbar_{s,k-1}, \nu_{s,k})\}$, and $g_{\epsilon^\mathsf{s}}(\hbar_{s,k-1}, \nu_{s,k}) := z -\min_z \{z,f_{\epsilon^\mathsf{s}}(\hbar_{s,k-1}, \nu_{s,k})\}$. Clearly, $g_{\epsilon^\mathsf{n}}(\hbar_{s,k-1}, \nu_{s,k})$ and $g_{\epsilon^\mathsf{s}}(\hbar_{s,k-1}, \nu_{s,k})$ both vanish at infinity for any fixed~$z$. It follows~that
\begin{align*}
	&\E_{\epsilon^\mathsf{n}} \Big[ \tilde{e}_{s,k}^T \tilde{e}_{s,k} \big| \mathbf{h}_{s,0:k-1} = 0 \Big] \\[1.5\jot]
	&=\int_{\mathbb{R}^m} \int_{\mathbb{R}^m}  f_{\epsilon^\mathsf{n}}(\hbar_{s,k-1}, \nu_{s,k})\Prob_{\epsilon^\mathsf{n}}( \hbar_{s,k-1} | \mathbf{h}_{s,0:k-1} = 0)\\[1.5\jot]
	&\qquad \qquad \times \Prob(\nu_{s,k}) d\hbar_{s,k-1} d\nu_{s,k}.
\end{align*}

In addition, we can write
\begin{align*}
	&\int_{\mathbb{R}^m} g_{\epsilon^\mathsf{n}}(\hbar_{s,k-1}, \nu_{s,k}) \Prob_{\epsilon^\mathsf{n}}( \hbar_{s,k-1} | \mathbf{h}_{s,0:k-2} = 0)\\[1\jot]
	&\qquad \times \Prob_{\epsilon^\mathsf{n}}(h_{s,k-1} = 0 | \hbar_{s,k-1}, \mathbf{h}_{s,0:k-2} = 0) d\hbar_{s,k-1} \\[1\jot]
	&\leq \int_{\mathbb{R}^m} g^*_{\epsilon^\mathsf{n}}(\hbar_{s,k-1}, \nu_{s,k}) \big(\Prob_{\epsilon^\mathsf{n}}( \hbar_{s,k-1} | \mathbf{h}_{s,0:k-2} = 0) \\[1\jot]
	&\qquad \times \Prob_{\epsilon^\mathsf{n}}(h_{s,k-1} = 0 | \hbar_{s,k-1}, \mathbf{h}_{s,0:k-2} = 0) \big)^* d\hbar_{s,k-1} \\[1\jot]
	&= \int_{\mathbb{R}^m} g_{\epsilon^\mathsf{s}}(\hbar_{s,k-1}, \nu_{s,k}) \big( \Prob_{\epsilon^\mathsf{n}}( \hbar_{s,k-1} | \mathbf{h}_{s,0:k-2} = 0) \\[1\jot]
	&\qquad \times \Prob_{\epsilon^\mathsf{n}}(h_{s,k-1} = 0 | \hbar_{s,k-1}, \mathbf{h}_{s,0:k-2} = 0) \big)^* d\hbar_{s,k-1} \\[1\jot]
	&\leq \int_{\mathbb{R}^m} g_{\epsilon^\mathsf{s}}(\hbar_{s,k-1}, \nu_{s,k}) \Prob_{\epsilon^\mathsf{s}}( \hbar_{s,k-1} | \mathbf{h}_{s,0:k-2} = 0) \\[1\jot]
	&\qquad \times \Prob_{\epsilon^\mathsf{s}}(h_{s,k-1} = 0 | \hbar_{s,k-1}, \mathbf{h}_{s,0:k-2} = 0) d\hbar_{s,k-1}
\end{align*}
where in the first inequality we used the Hardy-Littlewood inequality with respect to $\hbar_{s,k-1}$, in the equality the fact that $g^*_{\epsilon^\mathsf{n}}(\hbar_{s,k-1}, \nu_{,k}) = g_{\epsilon^\mathsf{s}}(\hbar_{s,k-1}, \nu_{s,k})$, and in the second inequality (\ref{eq:construction2B}) and Lemma~\ref{lemma:major}. This implies that
\begin{align*}
	&\int_{\mathbb{R}^m} \underset{z}{\min} \{z,f_{\epsilon^\mathsf{n}}(\hbar_{s,k-1}, \nu_{s,k}) \} \\
	&\qquad \qquad \times \Prob_{\epsilon^\mathsf{n}}(\hbar_{s,k-1} | \mathbf{h}_{s,0:k-1} = 0) d\hbar_{s,k-1} \nonumber\\[1\jot]
	&\geq \int_{\mathbb{R}^m} \underset{z}{\min} \{z,f_{\epsilon^\mathsf{s}}(\hbar_{s,k-1}, \nu_{s,k}) \}\\
	&\qquad \qquad \times \Prob_{\epsilon^\mathsf{s}}(\hbar_{s,k-1} | \mathbf{h}_{s,0:k-1} = 0) d\hbar_{s,k-1} .
\end{align*}
Taking $z$ to infinity in the above relation, we deduce that
\begin{align*}
	&\E_{\epsilon^\mathsf{n}} \Big[ \tilde{e}_{s,k}^T \tilde{e}_{s,k} \big| \mathbf{h}_{s,0:k-1} = 0 \Big]\\[1\jot]
	& =\int_{\mathbb{R}^m} \int_{\mathbb{R}^m}  f_{\epsilon^\mathsf{n}}(\hbar_{s,k-1}, \nu_{s,k}) \Prob_{\epsilon^\mathsf{n}}(\hbar_{s,k-1} | \mathbf{h}_{s,0:k-1} = 0)\\[1\jot]
	&\qquad \qquad \times \Prob( \nu_{s,k}) d \hbar_{s,k-1} d \nu_{s,k}\\[1\jot]
	& \geq \int_{\mathbb{R}^m} \int_{\mathbb{R}^m}  f_{\epsilon^\mathsf{s}}(\hbar_{s,k-1}, \nu_{s,k}) \Prob_{\epsilon^\mathsf{s}}(\hbar_{s,k-1} | \mathbf{h}_{s,0:k-1} = 0)\\[1\jot]
	&\qquad \qquad \times \Prob( \nu_{s,k}) d \hbar_{s,k-1} d \nu_{s,k}\\[1\jot]
	& = \E_{\epsilon^\mathsf{s}} \Big[ \tilde{e}_{s,k}^T \tilde{e}_{s,k} \big| \mathbf{h}_{s,0:k-1} = 0 \Big].
\end{align*}

In addition, for the terms involving the expected value of the cost-to-go, we find 
\begin{align*}
	&\E_{\epsilon^\mathsf{n}} \Big[ \Omega^{s,k+1,M}_{\epsilon^\mathsf{n}}(\tilde{e}_{s,k+1}) \big| \mathbf{h}_{s,0:k-1} = 0 , h_{s,k} = 1 \Big]\\[1\jot]
	& = \int_{\mathbb{R}^{m(k+2)}} \Omega^{s,k+1,M}_{\epsilon^\mathsf{n}}(\tilde{e}_{s,k+1})\\[1\jot]
	&\qquad \qquad \times \Prob_{\epsilon^\mathsf{n}}(\boldsymbol{\nu}_{s,0:k+1} | \mathbf{h}_{s,0:k-1} = 0, h_{s,k} = 1) d \boldsymbol{\nu}_{s, 0:k+1}.
\end{align*}
By Lemma~\ref{lem:mismatch-dyn-est}, when $h_{s,k}=1$, we have $\tilde{e}_{s,k+1} = K_{s,k+1} \nu_{s,k+1}$. Hence, $\tilde{e}_{s,k+1}$ is the same under both $\epsilon^\mathsf{n}$ and $\epsilon^\mathsf{s}$. For any $M$, let $\bar{\Omega}^{s,M}_{\epsilon^\mathsf{n}}(\tilde{e}_{s,0})$ represent a loss function that is structurally similar to $\Omega^{s,M}_{\epsilon^\mathsf{n}}(\tilde{e}_{s,0})$ but with new values of $\boldsymbol{\theta}_{s,0:M}$ and $\boldsymbol{\omega}_{s,0:M}$. Clearly, if $\Omega^{s,M}_{\epsilon^\mathsf{n}}(\tilde{e}_{s,0}) \geq \Omega^{s,M}_{\epsilon^\mathsf{s}}(\tilde{e}_{s,0})$ for any $\boldsymbol{\theta}_{s,0:M}$ and $\boldsymbol{\omega}_{s,0:M}$, then $\bar{\Omega}^{s,M}_{\epsilon^\mathsf{n}}(\tilde{e}_{s,0}) \geq \bar{\Omega}^{s,M}_{\epsilon^\mathsf{s}}(\tilde{e}_{s,0})$. 

Accordingly, we can write
\begin{align*}
	& \int_{\mathbb{R}^{m(k+2)}} \Omega^{s,k+1,M}_{\epsilon^\mathsf{n}} (\tilde{e}_{s,k+1}) \\[1\jot]
	&\qquad \times \Prob_{\epsilon^\mathsf{n}}(\boldsymbol{\nu}_{s,0:k+1} | \mathbf{h}_{s,0:k-1}=0, h_{s,k}=1) d \boldsymbol{\nu}_{s, 0:k+1} \\[1\jot]
	&= \int_{\mathbb{R}^m} \bar{\Omega}^{s,M-k-1}_{\epsilon^\mathsf{n}}(K_{s,k+1} \nu_{s,k+1}) \Prob(\nu_{s,k+1}) d \nu_{s, k+1}\\[1\jot]
	&\geq \int_{\mathbb{R}^m} \bar{\Omega}^{s,M-k-1}_{\epsilon^\mathsf{s}}(K_{s,k+1} \nu_{s,k+1}) \Prob(\nu_{s,k+1}) d \nu_{s, k+1}\\[1\jot]
	&= \int_{\mathbb{R}^{m(k+2)}} \Omega^{s,k+1,M}_{\epsilon^\mathsf{s}}(\tilde{e}_{s,k+1}) \\[1\jot]
	&\qquad \times \Prob_{\epsilon^\mathsf{s}}(\boldsymbol{\nu}_{s,0:k+1} | \mathbf{h}_{s,0:k-1}=0, h_{s,k}=1) d \boldsymbol{\nu}_{s, 0:k+1}
\end{align*}
where in the equalities we used the facts that $\Omega^{s,k+1,M}_{\epsilon^\mathsf{n}}(\tilde{e}) = \bar{\Omega}^{s,M-k-1}_{\epsilon^\mathsf{n}}(\tilde{e})$ for any Gaussian variable $\tilde{e}$ and a suitable selection of $\boldsymbol{\theta}_{s,0:M-k-1}$ and $\boldsymbol{\omega}_{s,0:M-k-1}$, and that $\nu_{s,k+1}$ is independent of $\mathbf{h}_{s,0:k}$, and adopted the Fubini's theorem; and in the inequality we used the hypothesis $\Omega^{s,M-k-1}_{\epsilon^\mathsf{n}}(\tilde{e}) \geq \Omega^{s,M-k-1}_{\epsilon^\mathsf{s}}(\tilde{e})$ for any Gaussian variable $\tilde{e}$. Therefore,
\begin{align*}
&\E_{\epsilon^\mathsf{n}} \Big[ \Omega^{s,k+1,M}_{\epsilon^\mathsf{n}}(\tilde{e}_{s,k+1}) \big| \mathbf{h}_{s,0:k-1} = 0 , h_{s,k} = 1\Big]\\[1.5\jot]
&\qquad \geq \E_{\epsilon^\mathsf{s}} \Big[ \Omega^{s,k+1,M}_{\epsilon^\mathsf{s}}(\tilde{e}_{s,k+1}) \big| \mathbf{h}_{s,0:k-1} = 0 , h_{s,k} = 1\Big].
\end{align*}
This establishes that $\Omega^{s,M}_{\epsilon^\mathsf{s}}(\tilde{e}_{s,0}) \leq \Omega^{s,M}_{\epsilon^\mathsf{n}}(\tilde{e}_{s,0})$, and verifies that $\Phi(\epsilon^\mathsf{s},\delta^\mathsf{o}) \leq \Phi(\epsilon^\mathsf{n},\delta^\mathsf{o})$.

\underline{Step 3.} We will show that $\Phi(\epsilon^\star,\delta^\star) \leq \Phi(\epsilon^\mathsf{s},\delta^\mathsf{o})$, where $\epsilon^\star$ is a special form of $\epsilon^\mathsf{s}$ and $\delta^\star$ is the same as $\delta^\mathsf{o}$. Observe that, by Lemmas~\ref{lem:estimator-decoder} and \ref{lem:zero-residuals}, when $\epsilon^\mathsf{s}$ is used, $\delta^\mathsf{o}$ must satisfy
\begin{align*}
	&\hat{x}_{s',k} =  A_{s',k-1} \hat{x}_{s',k-1} \nonumber \\[1.5\jot]
	&\qquad + u_{s',k-1} \gamma_{s',k-1} A_{s',k-1} (\check{x}_{s',k-1} - \hat{x}_{s',k-1})
\end{align*}
for $k \in \mathbb{N}_{[1,N]}$ and $s' \in \mathcal{M}$ with initial condition $\hat{x}_{s',0} = m_{s',0} = 0$. Moreover, by Lemmas 3 and 4, when $\epsilon^\mathsf{s}$ is used, we have
\begin{align}\label{eq:step3-etilde}
	\tilde{e}_{s',k} &= (1 - u_{s',k-1} \gamma_{s',k-1} ) \nonumber\\[1.5\jot]
	&\qquad \times A_{s',k-1} \tilde{e}_{s',k-1} + K_{s',k} \nu_{s',k}
\end{align}
for $k \in \mathbb{N}_{[1,N]}$ and $s' \in \mathcal{M}$ with initial condition $\tilde{e}_{s',0} = K_{s',0} \nu_{s',0}$. Hence, we can write
\begin{align*}
	&\E \Big[ \tilde{e}_{s',k+1}^T \tilde{e}_{s',k+1} \big| \mathcal{I}^{e_s}_k \Big]\\[1.5\jot]
	&\qquad = (1-u_{s',k} \lambda'_{s',k}) \tilde{e}_{s',k}^T A_{s',k}^T A_{s',k} \tilde{e}_{s',k}\nonumber\\[1.75\jot]
	&\qquad \quad + \tr(N_{s',k+1} K_{s',k+1}^T K_{s',k+1})
\end{align*}
where $N_{s',k+1}$ is the covariance of $\nu_{s',k+1}$.

For any $s \in \mathcal{M}$, we prove by induction that $V^{e_s}_{k}(\mathcal{I}^{e_s}_k)$ depends on $\tilde{e}_{s,k}$, $\tilde{e}_{\bar{s},k}$, $\lambda'_{s,k}$, and $\lambda'_{\bar{s},k}$, and is symmetric with respect to $\tilde{e}_{s,k}$ and $\tilde{e}_{\bar{s},k}$. Clearly, the claim holds for time $N$. We assume that the claim holds for time $k+1$ and we will show that it also holds for time $k$. For Problem 1, from the definition of the value function in (\ref{eq:Ve-def-bc}), we can write
\begin{align*}
	V^s_{k}(\mathcal{I}^{e_s}_k) &= \min_{u_{s,k} \in \{0, 1\}} \Big\{ \theta_{s,k} u_{s,k} \\[1\jot]
	&\qquad + \omega_{s,k+1} \E[ \tilde{e}_{s,k+1}^T \tilde{e}_{s,k+1} | \mathcal{I}^{e_s}_k] \\[1.75\jot]
	&\qquad + \omega_{\bar{s},k+1} \E[ \tilde{e}_{\bar{s},k+1}^T \tilde{e}_{\bar{s},k+1} | \mathcal{I}^{e_s}_k]  \\[1.75\jot]
	&\qquad + (\omega_{s,k+1}+ \omega_{\bar{s},k+1}) \tr Q_{s,k+1} \\[1.75\jot]
	&\qquad + \E[V^s_{k+1}(\mathcal{I}^{e_s}_{k+1})|\mathcal{I}^{e_s}_k] \Big\} \\[1\jot]
	&= \min_{u_{s,k} \in \{0, 1\}} \Big\{ \theta_{s,k} u_{s,k} \\[1.5\jot]
	&\qquad  + (1-u_{s,k} \lambda'_{s,k}) \omega_{s,k+1} \tilde{e}_{s,k}^T A_{s,k}^T A_{s,k} \tilde{e}_{s,k}\\[1\jot]
	&\qquad + (1-u_{s,k} \lambda'_{\bar{s},k}) \omega_{\bar{s},k+1} \tilde{e}_{\bar{s},k}^T A_{s,k}^T A_{s,k} \tilde{e}_{\bar{s},k}\\[1.75\jot]
	&\qquad + (\omega_{s,k+1}+ \omega_{\bar{s},k+1}) \tr(N_{s,k+1} K_{s,k+1}^T K_{s,k+1}) \\[1.75\jot]
	&\qquad + (\omega_{s,k+1}+ \omega_{\bar{s},k+1}) \tr Q_{s,k+1} \\[1.75\jot]
	&\qquad + \E[V^s_{k+1}(\mathcal{I}^{e_s}_{k+1})|\mathcal{I}^{e_s}_k] \Big\}
\end{align*}
for $k \in \mathbb{N}_{[0,N-1]}$ with initial condition $V^s_{N}(\mathcal{I}^{e_s}_{N}) = 0$, where we used the additivity property of $V^s_{k}(\mathcal{I}^{e_s}_k)$ and the fact that $\E[ \hat{e}_{i,k+1}^T \hat{e}_{i,k+1} | \mathcal{I}^{e_s}_k] = \E[ \tilde{e}_{i,k+1}^T \tilde{e}_{i,k+1} | \mathcal{I}^{e_s}_k] + \tr Q_{i,k+1}$ for $i \in \mathcal{M}$. Hence, $u^\star_{s,k} = 1$ if the following condition is satisfied
\begin{align*}
	&\theta_{s,k} + (1-\lambda'_{s,k}) \omega_{s,k+1} \tilde{e}_{s,k}^T A_{s,k}^T A_{s,k} \tilde{e}_{s,k}\\[1.75\jot]
	&\qquad \qquad + (1-\lambda'_{\bar{s},k}) \omega_{\bar{s},k+1} \tilde{e}_{\bar{s},k}^T A_{s,k}^T A_{s,k} \tilde{e}_{\bar{s},k}\\[1.75\jot]
	&\qquad \qquad +\E[V^s_{k+1}(\mathcal{I}^{e_s}_{k+1})|\mathcal{I}^{e_s}_k, u_{s,k} = 1]\\[1.75\jot]
	&\leq \omega_{s,k+1} \tilde{e}_{s,k}^T A_{s,k}^T A_{s,k} \tilde{e}_{s,k}\\[1.75\jot]
	&\qquad \qquad + \omega_{\bar{s},k+1}\tilde{e}_{\bar{s},k}^T A_{s,k}^T A_{s,k} \tilde{e}_{\bar{s},k}\\[1.75\jot]
	&\qquad \qquad +\E[V^s_{k+1}(\mathcal{I}^{e_s}_{k+1})|\mathcal{I}^{e_s}_k, u_{s,k} = 0]
\end{align*} 
and $u^\star_{s,k} = 0$ otherwise. Therefore, we can conclude that (13) holds.

However, for Problem 2, from the definition of the value function in (\ref{eq:Ve-def-mac}), we can write
\begin{align*}
	V^s_{k}(\mathcal{I}^{e_s}_k) &= \min_{u_{s,k} \in \{0, 1\} : u_{\bar{s},k} = u^\star_{\bar{s},k} } \Big\{\theta_{s,k} u_{s,k} + \theta_{\bar{s},k} u_{\bar{s},k} \\[1.75\jot]
	&\qquad + \omega_{s,k+1} \E[ \tilde{e}_{s,k+1}^T \tilde{e}_{s,k+1} | \mathcal{I}^{e_s}_k] \\[1.75\jot]
	&\qquad + \omega_{\bar{s},k+1} \E[ \tilde{e}_{\bar{s},k+1}^T \tilde{e}_{\bar{s},k+1} | \mathcal{I}^{e_s}_k]  \\[1.75\jot]
	&\qquad + \omega_{s,k+1} \tr Q_{s,k+1} + \omega_{\bar{s},k+1} \tr Q_{\bar{s},k+1} \\[1.75\jot]
	&\qquad + \E[V^s_{k+1}(\mathcal{I}^{e_s}_{k+1})|\mathcal{I}^{e_s}_k] \Big\}\\[1\jot]
	&= \min_{u_{s,k} \in \{0, 1\} : u_{\bar{s},k} = u^\star_{\bar{s},k} } \Big\{\theta_{s,k} u_{s,k} + \theta_{\bar{s},k} u_{\bar{s},k} \\[1.75\jot]
	&\qquad + (1-u_{s,k} \lambda'_{s,k}) \omega_{s,k+1} \tilde{e}_{s,k}^T A_{s,k}^T A_{s,k} \tilde{e}_{s,k}	\\[1.75\jot]
	&\qquad + (1-u_{\bar{s},k} \lambda'_{\bar{s},k}) \omega_{\bar{s},k+1} \tilde{e}_{\bar{s},k}^T A_{\bar{s},k}^T A_{\bar{s},k} \tilde{e}_{\bar{s},k}\\[1.75\jot]
	&\qquad + \omega_{s,k+1} \tr(N_{s,k+1} K_{s,k+1}^T K_{s,k+1})\\[1.75\jot]
	&\qquad + \omega_{\bar{s},k+1} \tr(N_{\bar{s},k+1} K_{\bar{s},k+1}^T K_{\bar{s},k+1}) \\[1.75\jot]
	&\qquad + \omega_{s,k+1} \tr Q_{s,k+1} + \omega_{\bar{s},k+1} \tr Q_{\bar{s},k+1} \nonumber\\[1\jot]
	&\qquad + \E[V^s_{k+1}(\mathcal{I}^{e_s}_{k+1})|\mathcal{I}^{e_s}_k] \Big\}
\end{align*}
subject to $u_{s,k} + u_{\bar{s},k} \leq 1$, for $k \in \mathbb{N}_{[0,N-1]}$ with initial condition $V^s_{N}(\mathcal{I}^{e_s}_{N}) = 0$, where we used the additivity property of $V^s_{k}(\mathcal{I}^{e_s}_k)$ and the fact that $\E[ \hat{e}_{j,k+1}^T \hat{e}_{j,k+1} | \mathcal{I}^{e_s}_k] = \E[ \tilde{e}_{j,k+1}^T \tilde{e}_{j,k+1} | \mathcal{I}^{e_s}_k] + \tr Q_{j,k+1}$ for $j \in \mathcal{M}$. Hence, $u^\star_{s,k} = 1$ if the following conditions are satisfied
\begin{align*}
	&\theta_{s,k} + (1-\lambda'_{s,k}) \omega_{s,k+1} \tilde{e}_{s,k}^T A_{s,k}^T A_{s,k} \tilde{e}_{s,k}\\[1.75\jot]
	&\qquad \qquad +\E[V^s_{k+1}(\mathcal{I}^{e_s}_{k+1})|\mathcal{I}^{e_s}_k, u_{s,k} = 1, u_{\bar{s},k} = 0]\\[1.75\jot]
	&\leq \omega_{s,k+1} \tilde{e}_{s,k}^T A_{s,k}^T A_{s,k} \tilde{e}_{s,k}\\[1.75\jot]
	&\qquad \qquad +\E[V^s_{k+1}(\mathcal{I}^{e_s}_{k+1})|\mathcal{I}^{e_s}_k, u_{s,k} = 0, u_{\bar{s},k} = 0]
\end{align*}
and
\begin{align*}	
	&\theta_{s,k} + (1-\lambda'_{s,k}) \omega_{s,k+1} \tilde{e}_{s,k}^T A_{s,k}^T A_{s,k} \tilde{e}_{s,k}\\[1.75\jot]
	&\qquad \qquad + \omega_{\bar{s},k+1} \tilde{e}_{\bar{s},k}^T A_{\bar{s},k}^T A_{\bar{s},k} \tilde{e}_{\bar{s},k}\\[1.75\jot]
	&\qquad \qquad +\E[V^s_{k+1}(\mathcal{I}^{e_s}_{k+1})|\mathcal{I}^{e_s}_k, u_{s,k} = 1, u_{\bar{s},k} = 0]
\end{align*} 
\begin{align*}
	&\leq \theta_{\bar{s},k} + (1-\lambda'_{\bar{s},k}) \omega_{\bar{s},k+1} \tilde{e}_{\bar{s},k}^T A_{\bar{s},k}^T A_{\bar{s},k} \tilde{e}_{\bar{s},k}\\[1.75\jot]
	&\qquad \qquad + \omega_{s,k+1} \tilde{e}_{s,k}^T A_{s,k}^T A_{s,k} \tilde{e}_{s,k}\\[1.75\jot]
	&\qquad \qquad +\E[V^s_{k+1}(\mathcal{I}^{e_s}_{k+1})|\mathcal{I}^{e_s}_k, u_{s,k} = 0, u_{\bar{s},k} = 1]
\end{align*} 
and $u^\star_{s,k} = 0$ otherwise. Since the same set of conditions can be written for $u^\star_{\bar{s},k}$, we can conclude that (18) holds. 

Finally, note that, by the hypothesis, $V^{e_s}_{k+1}(\mathcal{I}^{e_s}_{k+1})$ depends on $\tilde{e}_{s,k+1}$, $\tilde{e}_{\bar{s},k+1}$, $\lambda'_{s,k+1}$, and $\lambda'_{\bar{s},k+1}$, and is symmetric with respect to $\tilde{e}_{s,k+1}$ and $\tilde{e}_{\bar{s},k+1}$. Plugging (\ref{eq:step3-etilde}) into $V^{e_s}_{k+1}(\mathcal{I}^{e_s}_{k+1})$, we can calculate $\E[ V^{e_s}_{k+1}(\mathcal{I}^{e_s}_{k+1}) | \mathcal{I}^{e_s}_k,u_{s,k}]$ while $\nu_{s,k+1}$, $\nu_{\bar{s},k+1}$, $\gamma_{s,k}$, and $\gamma_{\bar{s},k}$ are averaged out. Moreover, following the facts that $\nu_{s,k+1}$ and $\nu_{\bar{s},k+1}$ are Gaussian variables with zero mean and that $\lambda'_{s,k+1}$ and $\lambda'_{\bar{s},k+1}$ depend only on $\lambda'_{s,k}$ and $\lambda'_{\bar{s},k}$, respectively, we can deduce that $\E[ V^{e_s}_{k+1}(\mathcal{I}^{e_s}_{k+1}) | \mathcal{I}^{e_s}_k,u_{s,k}]$ depends on $\tilde{e}_{s,k}$, $\tilde{e}_{\bar{s},k}$, $\lambda'_{s,k}$, and $\lambda'_{\bar{s},k}$, and is symmetric with respect to $\tilde{e}_{s,k}$ and $\tilde{e}_{\bar{s},k}$. This proves the claim in this step, and completes the proof.
\end{IEEEproof}

Next, we provide the proof of Corollaries 1 and 2.
\begin{IEEEproof}
The proof is based on a direct application of Theorems 1 and 2. We only need to determine $u_{s,0}$ and $\hat{x}_{s,1}$ for $s \in \mathcal{M}$. Note that, when $N = 1$, we have $\Delta_{c,0} = \Delta^I_{j,0} = \Delta^{II}_{j,0} = 0$ for $j \in \mathcal{M}$. Moreover, we know that $\tilde{e}_{s,0} = K_{s,0} \nu_{s,0}$. We obtain the results by incorporating these terms in the results of Theorems 1 and 2.
\end{IEEEproof}

\section{Numerical Example}\label{sec:example}
In this section, we provide a numerical example pertaining to satellite communications to demonstrate how the framework developed in the previous sections can be used for state estimation over broadcast and multi-access channels. Our example is based on spin-stabilized spacecraft. In spin-stabilized spacecraft, the body is spinning about the $z$-axis, i.e., the axis of symmetry, with angular velocity $\omega_z$. Let $\omega_z = \omega_0$ be constant. Then, the Euler equation is written as
\begin{align*}\setlength\arraycolsep{6pt}\def\arraystretch{1.5}
\begin{bmatrix}
 \dot{\omega}_x \\
 \dot{\omega}_y \\
 \dot{\omega}_z
 \end{bmatrix}
 =
\begin{bmatrix}
 0 & \frac{I_y - I_z}{I_x} \omega_0 & 0 \\
 \frac{I_z - I_x}{I_y} \omega_0 & 0 & 0 \\
 0 & 0 & 0
 \end{bmatrix}
\begin{bmatrix}
 \omega_x \\
 \omega_y \\
 \omega_z
 \end{bmatrix} 
+
\begin{bmatrix}
e_x \\
e_y \\
e_z
\end{bmatrix}
\end{align*}
where $(I_x,I_y,I_z)$ is the moment of inertia and $(e_x,e_y,e_z)$ is a Gaussian disturbance torque. Note that for spin stability, the spin axis must be either the major or minor axis of inertia. In this example, we use $\omega_0 = 2 \pi \ \text{rad/s}$, $I_x = I_y = 20 \ \text{kg.m$^2$}$, $I_z = 100 \ \text{kg.m$^2$}$, and discretize the Euler equation over the time horizon $N = 1000$. Suppose that there are two spacecraft and two ground stations. Each spacecraft is equipped with a sensor that partially observes each component of the angular velocity at each time $k$. The state and output equations can be expressed by $A_{s,k} = [0.4258, 0.4258, 0; 0.4258, 0.4258, 0; 0, 0 , 1]$, $W_{s,k} = 10^{-6} \diag\{0.2245,0.2245,0.0025\}$, $C_{s,k} = \diag \{1,1,1\}$, and $V_{s,k} = 10^{-3} \diag \{ 1, 1, 1\}$, for $s \in \{1,2\}$ and $k \in \mathbb{N}_{[0,N]}$.

\begin{figure}[t]
\centering
  \includegraphics[width=.94\linewidth]{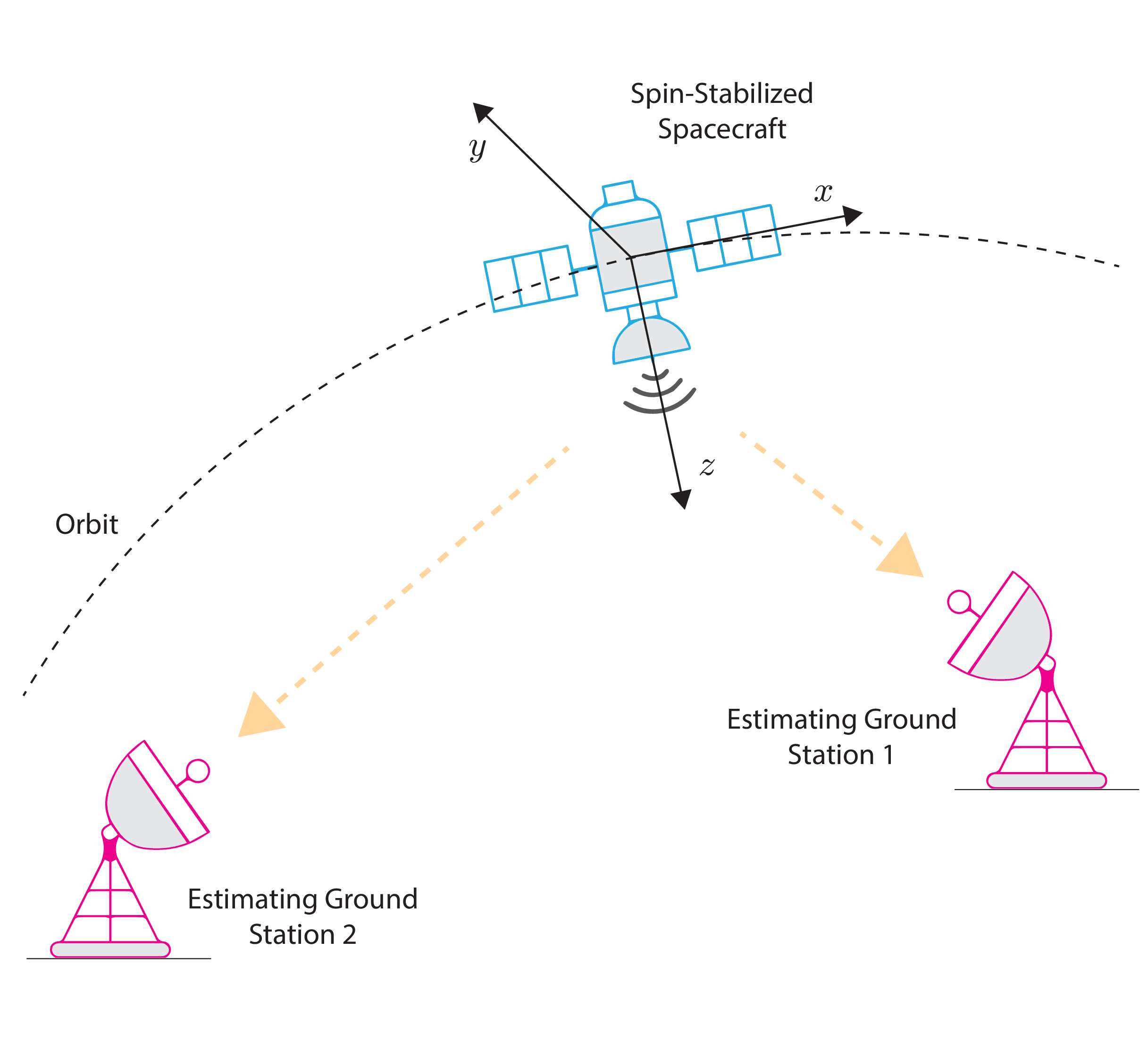}
  \caption{Angular velocity estimation of a perturbed spin-stabilized spacecraft over a packet-erasure broadcast channel at two ground station. The objective is to find optimal encoding and decoding strategies.}
  \label{fig:satellite-bc}
\end{figure}

\begin{figure}[t!]
\centering
  \includegraphics[width=.94\linewidth]{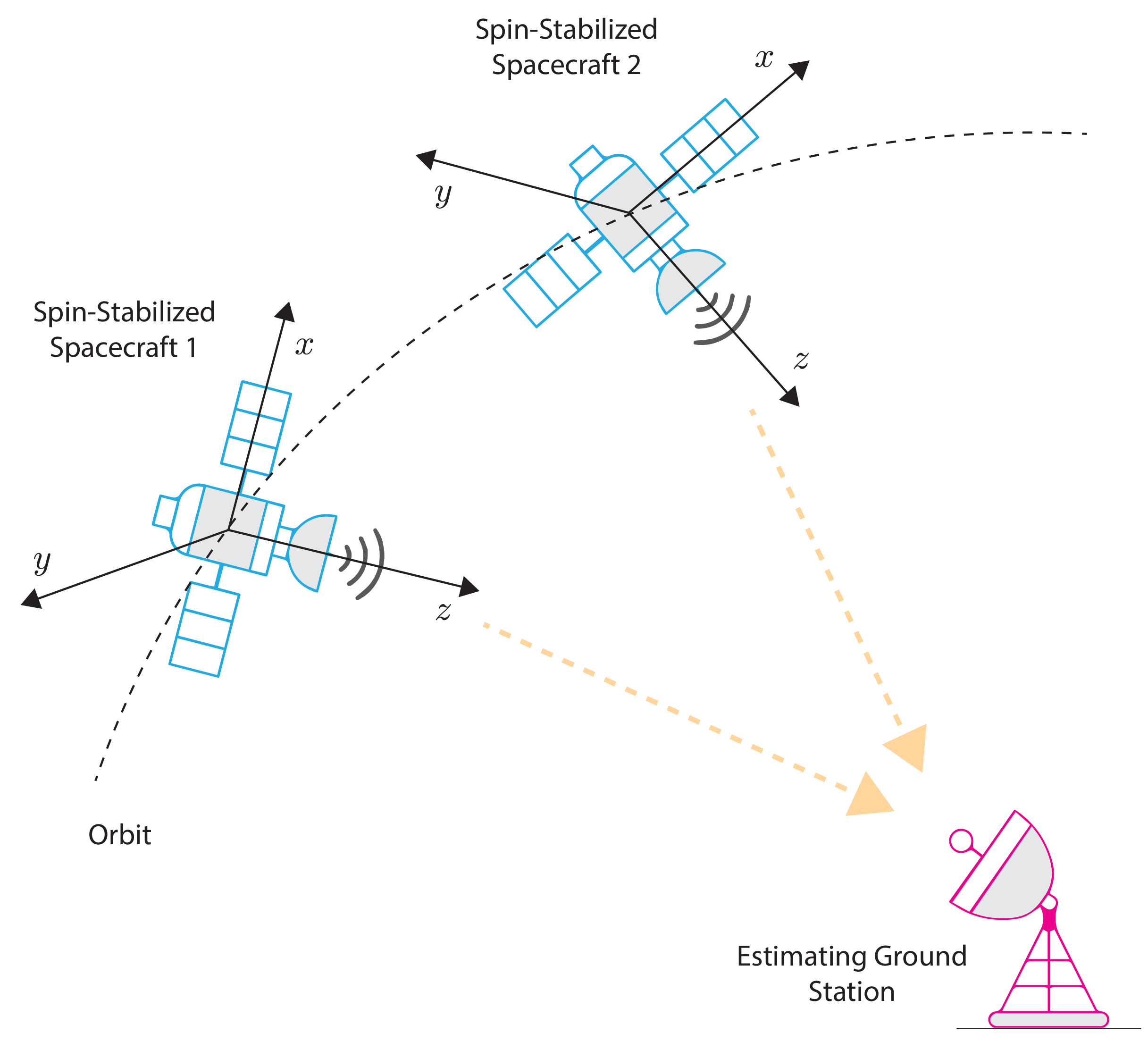}
  \caption{Angular velocity estimation of two perturbed spin-stabilized spacecraft over a packet-erasure multi-access channel at a ground station. The objective is to find optimal encoding and decoding strategies.}
  \label{fig:satellite-mac}
\end{figure}

In the first scenario, the measurements of a spacecraft should be transmitted over a broadcast channel to ground stations, where the angular velocity of the spacecraft is estimated (see Fig.~\ref{fig:satellite-bc}). The broadcast channel is subject to packet loss with $\lambda_{1,k} = 0.3$ and $\lambda_{2,k} = 0.1$ for all $k \in \mathbb{N}_{[0,N]}$, and with one-step time delay. In the second scenario, the measurements of the two spacecraft should be transmitted over a multi-access channel to a ground station, where the angular velocities of the spacecraft are estimated (see Fig.~\ref{fig:satellite-mac}). The multi-access channel is subject to packet loss with $\lambda_{1,k} = 0.3$ and $\lambda_{2,k} = 0.1$ for all $k \in \mathbb{N}_{[0,N]}$, and with one-step time delay. In both scenarios, we are interested in finding the optimal encoding and decoding policies in the the sense of the frequency-distortion tradeoff, with weighting coefficients $\theta_{c,k} = 1.1 \times 10^{-5}$, $\theta_{1,k} = \theta_{2,k} = 0.5 \times 10^{-5}$, and $\omega_{1,k} = \omega_{2,k} = 1$ for $k\in\mathbb{N}_{[0,N]}$.

\begin{figure*}[t!]
\centering
  \includegraphics[width=.98\linewidth]{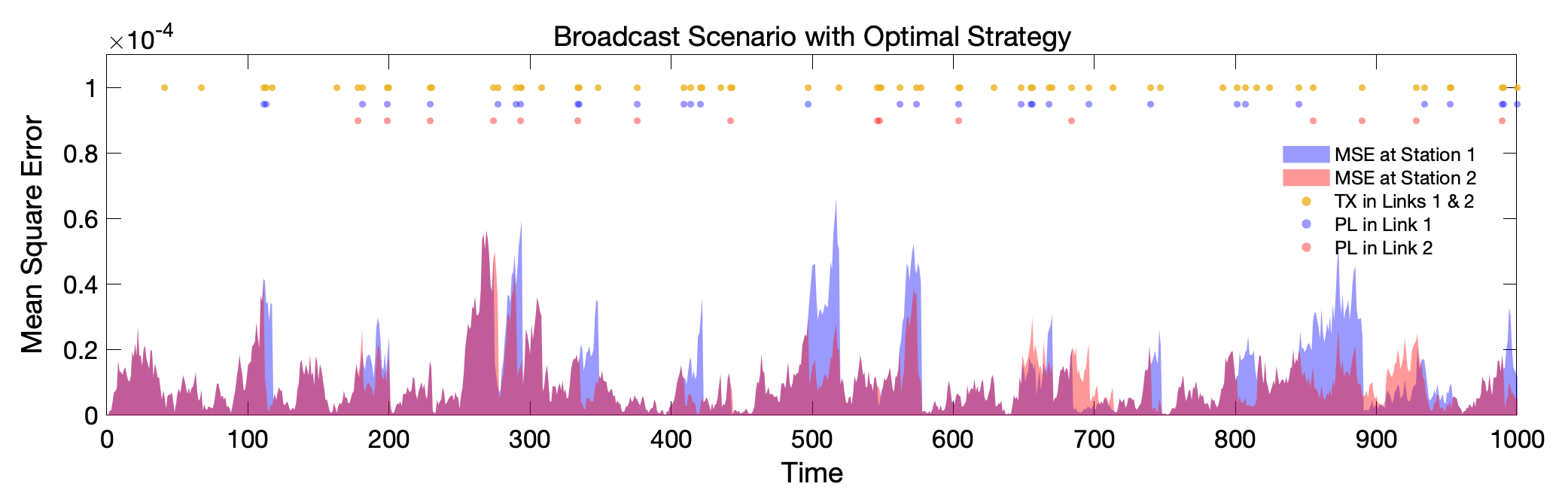}
  \caption{MSE~and packet transmission trajectories when the optimal broadcast policy is adopted. In this experiment, the total MSE at station~1 is $0.0126$ and at station~2 is $0.0101$; the total number of transmissions (TX) in both links is $65$; and the total number of packet losses (PL) in link~1 is $32$ and in link~2 is $16$.}
  \label{fig:mse-broadcast}
\end{figure*}

\begin{figure*}[t!]
\centering
  \includegraphics[width=.98\linewidth]{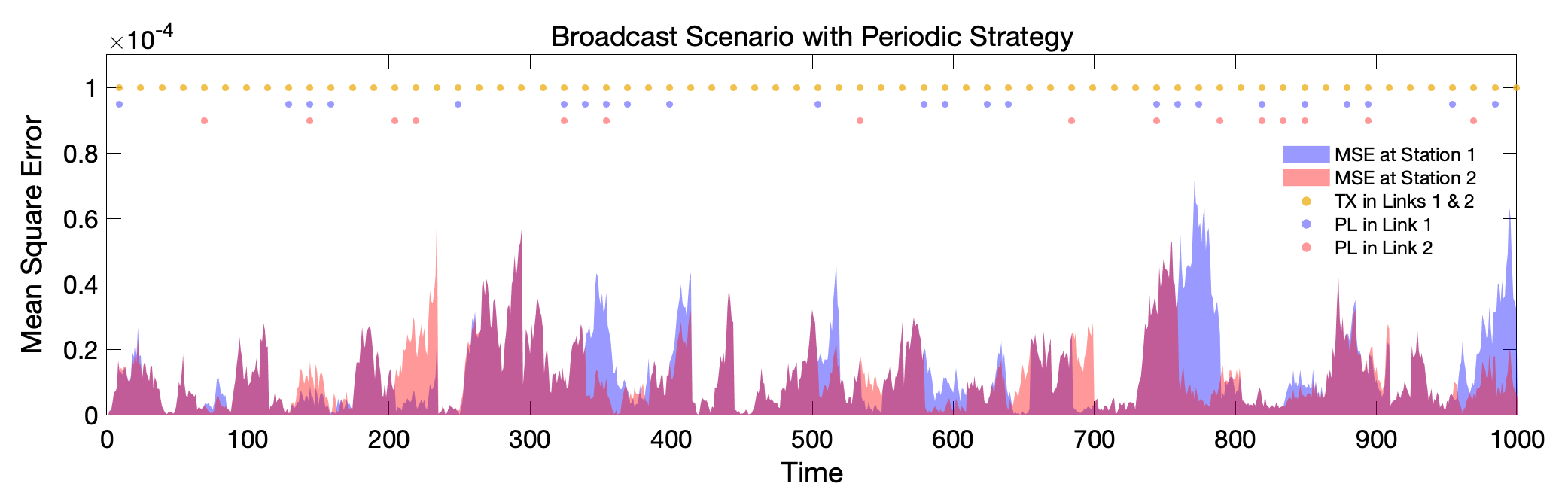}
  \caption{MSE~and packet transmission trajectories when a periodic broadcast policy is adopted. In this experiment, the total MSE at station~1 is $0.0142$ and at station~2 is $0.0123$; the total number of transmissions (TX) in both links is $67$; and the total number of packet losses (PL) in link~1 is $24$ and in link~2 is $15$.}
  \label{fig:mse-broadcast-p}
\end{figure*}

\begin{figure*}[t!]
\centering
  \includegraphics[width=.98\linewidth]{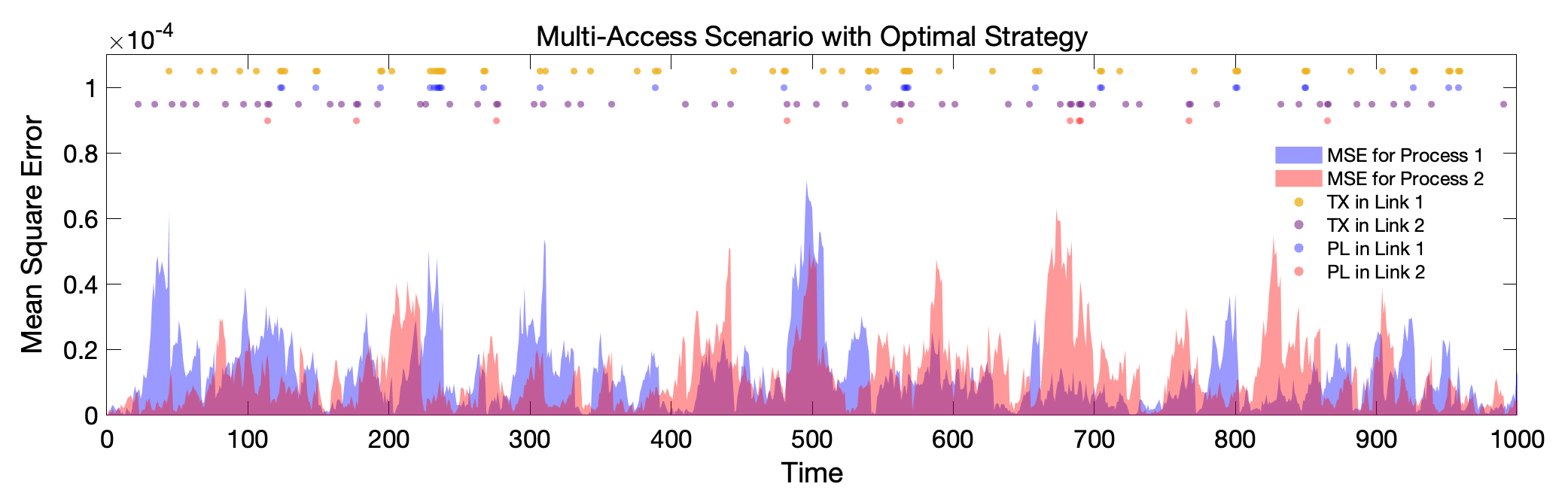}
  \caption{MSE and packet transmission trajectories when the optimal multi-access policy is adopted. In this experiment, the total MSE associated with process~1 is $0.0127$ and with process~2 is $0.0128$; the total number of transmissions (TX) in each link is $65$; and the total number of packet losses (PL) in link~1 is $29$ and in link~2 is $10$.}
  \label{fig:mse-multiaccess}
\end{figure*}

\begin{figure*}[t!]
\centering
  \includegraphics[width=.98\linewidth]{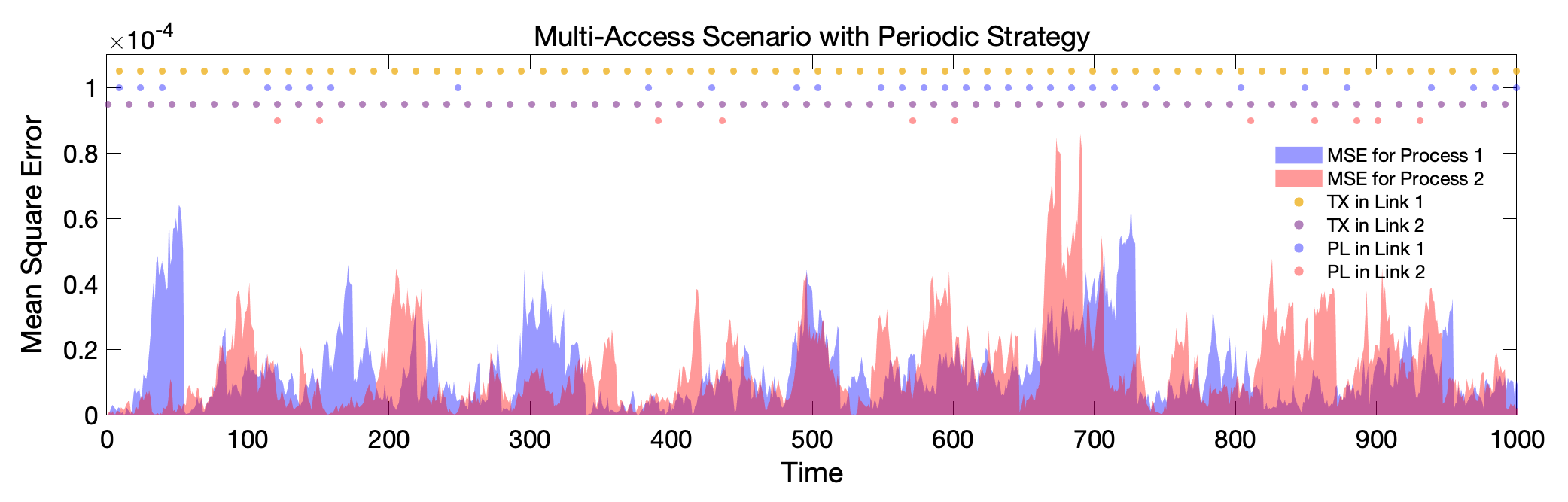}
  \caption{MSE and packet transmission trajectories when a periodic multi-access policy is adopted. In this experiment, the total MSE associated with process~1 is $0.0140$ and with process~2 is $0.0151$; the total number of transmissions (TX) in each link is $67$; and the total number of packet losses (PL) in link~1 is $32$ and in link~2 is $11$.}
  \label{fig:mse-multiaccess-p}
\end{figure*}

In the broadcast scenario, the MSE~and packet transmission trajectories for a simulated realization are shown in Figs.~\ref{fig:mse-broadcast}~and~\ref{fig:mse-broadcast-p}. More specifically, when the optimal broadcast policy is adopted, the total MSE~at station~1 is $0.0126$ and at station~2 is $0.0101$, the total number of simultaneous transmissions in both links is $65$, and the total number of packet losses in link~1 is $32$ and in link~2 is $16$. However, when a periodic policy is adopted, the total MSE~at station~1 is $0.0142$ and at station~2 is $0.0123$, the total number of simultaneous transmissions in both links is $67$, and the total number of packet losses in link~1 is $24$ and in link~2 is $15$. In the multi-access scenario, the MSE~and packet transmission trajectories for a simulated realization are shown in Figs.~\ref{fig:mse-multiaccess}~and~\ref{fig:mse-multiaccess-p}. More specifically, when the optimal multi-access policy is adopted, the total MSE~associated with process~1 is $0.0127$ and with process~2 is $0.0128$, the total number of transmissions in each link is $65$, and the total number of packet losses in link~1 is $29$ and in link~2 is $10$. However, when a periodic multi-access policy is adopted, the total MSE~associated with process~1 is $0.0140$ and with process~2 is $0.0151$, the total number of transmissions in each link is $67$, and the total number of packet losses in link~1 is $32$ and in link~2 is $11$. We observe in this example that the optimal broadcast and multi-access policies proved effective in improving the system performance. It is interesting to note that, in comparison with the periodic policies, the optimal scheduling policies not only transmit sensory information less frequently when the estimation discrepancy is small, but transmit more frequently and more persistently when the estimation discrepancy is large and when there have been some recent packet losses.

\section{Conclusions}\label{sec:conclusions}
This article examined the problem of state estimation over multi-terminal channels in an unreliable regime. We focused on two canonical settings. In the first setting, measurements of a common stochastic source need to be transmitted to two distinct remote monitors over a packet-erasure broadcast channel. In the second setting, measurements of two distinct stochastic sources need to be transmitted to a common remote monitor over a packet-erasure multi-access channel. For these networked systems, we identified optimal scheduling and estimation strategies in the sense of a causal tradeoff between the estimation error and the communication cost. These strategies unveil the fundamental performance limits of the underlying networked systems. Future research should extend the results of this article to a broadcast setting with private information for the receivers, and a multi-access setting without coordination between the transmitters.

\section{Appendix}
In this section, we present auxiliary results that are exploited for the derivation of the main results. The next two lemmas characterize the optimal estimators at the encoders and the decoders.

\begin{lemma}\label{lem:estimator-encoder}\emph{
The optimal estimators minimizing the MSE~at the encoders satisfy
\begin{align}
	\check{x}_{s,k} &= m_{s,k} + K_{s,k} ( y_{s,k} - C_{s,k} m_{s,k} ) \label{eq:est-KF-xhat} \\[1.5\jot]
	m_{s,k} &= A_{s,k-1} \check{x}_{s,k-1} \label{eq:est-KF-m}\\[1.5\jot]
	Q_{s,k} &= ( M_{s,k}^{-1} + C_{s,k}^T V_{s,k}^{-1} C_{s,k} )^{-1}\\[1.5\jot]
	M_{s,k} &= A_{s,k-1} Q_{s,k-1} A_{s,k-1}^T + W_{s,k-1}
\end{align}
for $k \in \mathbb{N}_{[1,N]}$ and $s \in \mathcal{M}$ with initial conditions $\check{x}_{s,0} = m_{s,0} + K_{s,0}(y_{s,0} - C_{s,0} m_{s,0})$ and $Q_{s,0} = (M_{s,0}^{-1} + C_{s,0}^T V_{s,0}^{-1} C_{s,0})^{-1}$, where $m_{s,k} = \E[ x_{s,k} | \mathcal{I}^{e_s}_{k-1}]$, $K_{s,k} = Q_{s,k} C_{s,k}^T V_{s,k}^{-1}$, $Q_{s,k} = \Cov[x_{s,k} | \mathcal{I}^{e_s}_{k}]$, and $M_{s,k} = \Cov[x_{s,k} | \mathcal{I}^{e_s}_{k-1}]$.}
\end{lemma}

\begin{IEEEproof}
Observe that, given $\mathcal{I}^{e_s}_k$ at the $s$th encoder, the MMSE estimator is $\E[ x_{s,k} | \mathcal{I}^{e_s}_k]$. This estimator must satisfy the Kalman filter equations (see, e.g., \cite{stengel1994}). 
\end{IEEEproof}

\begin{lemma}\label{lem:estimator-decoder}\emph{
The optimal estimators minimizing the MSE~at the decoders satisfy
\begin{align}\label{eq:est-monitor}
	\hat{x}_{s,k} &=  A_{s,k-1} \hat{x}_{s,k-1} \nonumber\\[1.5\jot]
	&\quad + u_{s,k-1} \gamma_{s,k-1} A_{s,k-1} (\check{x}_{s,k-1} - \hat{x}_{s,k-1}) \nonumber\\[1.5\jot]
	&\quad + (1-u_{s,k-1}\gamma_{s,k-1}) \imath_{s,k-1}
\end{align}
for $k \in \mathbb{N}_{[1,N]}$ and $s \in \mathcal{M}$ with initial condition $\hat{x}_{s,0} = m_{s,0}$, where $\imath_{s,k-1} = A_{s,k-1} \E[\hat{e}_{s,k-1} | \mathcal{I}_{k-1}^{d_s}, u_{s,k-1} \gamma_{s,k-1} = 0]$.}
\end{lemma}

\begin{IEEEproof}
Observe that, given $\mathcal{I}^{d_s}_k$ at the $s$th decoder, the MMSE estimator is $\E[ x_{s,k} | \mathcal{I}^{d_s}_k]$. Taking the conditional expectation of the state equation given $\mathcal{I}^{d_s}_k$, we obtain
\begin{align}\label{eq:propagation}
\E \Big[ x_{s,k} \big| \mathcal{I}^{d_s}_k \Big] = A_{s,k-1} \E \Big[x_{s,k-1} \big| \mathcal{I}_k^{d_s} \Big]
\end{align}
for $k \in \mathbb{N}_{[1,N]}$ as $w_{s,k-1}$ is independent of $\mathcal{I}^{d_s}_k$ and has zero mean. Suppose that the index of the transmitted measurement is known at the decoder. When measurement is successfully delivered at time $k$, we have $z_{s,k} = \check{x}_{s,k-1}$. In this case, we get $\E[x_{s,k-1} | \mathcal{I}^{d_s}_k] = \E[x_{s,k-1} | \check{x}_{s,k-1}, Q_{s,k-1}] = \check{x}_{s,k-1}$ as $\{ \check{x}_{s,k-1}, Q_{s,k-1} \}$ is a sufficient statistic of $\mathcal{I}^{d_s}_{k}$ with respect to $x_{s,k-1}$. Hence, using (\ref{eq:propagation}), when $z_{s,k} = \check{x}_{s,k-1}$, i.e., when $u_{s,k-1} \gamma_{s,k-1} = 1$, we~get
\begin{align}\label{eq:update0}
	\E \Big[x_{s,k} \big| \mathcal{I}^{d_s}_k \Big] = A_{s,k-1} \check{x}_{s,k-1}
\end{align}
for $k \in \mathbb{N}_{[1,N]}$. However, when no measurement is successfully delivered at time $k$, we have $z_{s,k} = \mathfrak{E}$. Define $l_{s,k-1} := \E[x_{s,k-1} | \mathcal{I}^{d_s}_{k-1}, u_{s, k-1} \gamma_{s,k-1} = 0] - \E[x_{s, k-1} | \mathcal{I}^{d_s}_{k-1}]$ when $z_{s,k} = \mathfrak{E}$. Then, using (\ref{eq:propagation}) and the definition of $l_{s,k-1}$, when $z_{s,k} = \mathfrak{E}$, i.e., when $u_{s,k-1} \gamma_{s,k-1} = 0$, we get
\begin{align}\label{eq:erasure-update1}
	\E \Big[x_{s,k} \big| \mathcal{I}^{d_s}_k \Big] =  A_{s,k-1} \hat{x}_{s,k-1} + A_{s,k-1} l_{s,k-1}
\end{align}
for $k \in \mathbb{N}_{[1,N]}$, where we used the fact that $\{\mathcal{I}^{d_s}_{k-1}, z_{s,k} = \mathfrak{E}\}$ is equivalent to $\mathcal{I}^{d_s}_k$. Now, define $\imath_{s,k-1} := A_{s,k-1} l_{s,k-1}$ as the signaling residual. We can obtain (\ref{eq:est-monitor}) by combining (\ref{eq:update0}) and (\ref{eq:erasure-update1}). Finally, the initial condition is $\E[x_{s,0}] = m_{s,0}$, as no measurement is available at the decoder at time $k = 0$.
\end{IEEEproof}

The next two lemmas provide certain properties pertaining to the estimation mismatches and the signaling residuals.

\begin{lemma}\label{lem:mismatch-dyn-est}\emph{
The estimation mismatches satisfy
\begin{align}\label{eq:et-dynamics}
	\tilde{e}_{s,k} &= (1 - u_{s,k-1} \gamma_{s,k-1} ) A_{s,k-1} \tilde{e}_{s,k-1} \nonumber\\[1.5\jot]
	&+ K_{s,k} \nu_{s,k} - (1-u_{s,k-1} \gamma_{s,k-1}) \imath_{s,k-1}
\end{align}
for $k \in \mathbb{N}_{[1,N]}$ and $s \in \mathcal{M}$ with initial condition $\tilde{e}_{s,0} = K_{s,0} \nu_{s,0}$.}
\end{lemma}

\begin{IEEEproof}
Observe that we can obtain (\ref{eq:et-dynamics}) for $s \in \mathcal{M}$ by plugging (\ref{eq:est-KF-m}) into (\ref{eq:est-KF-xhat}), and then subtracting (\ref{eq:est-monitor}) from the result. 
\end{IEEEproof}

\begin{lemma}\label{lem:zero-residuals}\emph{
Let $\Prob(u_{s,k} | \boldsymbol{\nu}_{s,0:k}, \mathbf{u}_{s,0:k-1})$ be a symmetric function with respect to $\boldsymbol{\nu}_{s,0:k}$ for $k \in \mathbb{N}_{[0,N]}$ and $s \in \mathcal{M}$. Then, $\imath_{s,k}$ in Lemmas~\ref{lem:estimator-decoder} and \ref{lem:mismatch-dyn-est} are equal to zero for $k \in \mathbb{N}_{[0,N-1]}$ and $s \in \mathcal{M}$.}
\end{lemma}

\begin{IEEEproof}
Observe that $\tilde{e}_{s,0} = K_{s,0} \nu_{s,0}$. We assume that $\imath_{s,t} = 0$ for all $t \in \mathbb{N}_{[0,k-1]}$, and will show that $\imath_{s,k} = 0$. We can express $\Prob ( \mathbf{u}_{s,0:k} \big| \boldsymbol{\nu}_{s,0:k})$ based on the following decomposition $\Prob ( \mathbf{u}_{s,0:k} \big| \boldsymbol{\nu}_{s,0:k}) = \prod_{t=0}^{k} \Prob (u_{s,t} \big| \boldsymbol{\nu}_{s,0:t}, \mathbf{u}_{s,0:t-1})$, where we used the fact that $u_{s,t}$ is independent of $\boldsymbol{\nu}_{s,t+1:k}$. Therefore, by the hypothesis, $\Prob( \mathbf{u}_{s,0:k} | \boldsymbol{\nu}_{s,0:k})$ is a symmetric function with respect to $\boldsymbol{\nu}_{s,0:k}$. In addition, by Bayes' theorem, we have $\Prob (\boldsymbol{\nu}_{s,0:k} | \mathbf{u}_{s,0:k}) \propto \Prob ( \mathbf{u}_{s,0:k} | \boldsymbol{\nu}_{s,0:k} ) \Prob (\boldsymbol{\nu}_{s,0:k})$. As $\Prob(\boldsymbol{\nu}_{s,0:k})$ is a symmetric distribution, we deduce that $\Prob(\boldsymbol{\nu}_{s,0:k} | \mathbf{u}_{s,0:k})$ is also a symmetric distribution with respect to $\boldsymbol{\nu}_{s,0:k}$. Finally, by marginalization and the fact that $\boldsymbol{\nu}_{s, k':k}$ is independent of $\mathbf{u}_{s,0:k'-1}$ for any $k' \in \mathbb{N}_{[0,k]}$, we find that $\Prob(\boldsymbol{\nu}_{s,k':k} | \mathbf{u}_{s,k':k})$ is a symmetric distribution with respect to $\boldsymbol{\nu}_{s,k':k}$ as well.

Now, using the tower properties of conditional expectations and the fact that $\gamma_{s,k}$ provides no information about $\hat{e}_{s,k}$, we observe that $\E[ \hat{e}_{s,k} | \mathcal{I}^{d_s}_k, u_{s,k} \gamma_{s,k}] = \E[ \tilde{e}_{s,k} | \mathcal{I}^{d_s}_k, u_{s,k} \gamma_{s,k}]$. In addition, by (\ref{eq:et-dynamics}), we obtain
\begin{align}\label{eq:lem6:et-dyn}
	\tilde{e}_{s,k} &= \textstyle \sum_{t=1}^{\eta_{s,k}+1} \Big(\prod_{t'=1}^{t-1} A_{s,k-t'} \Big) K_{s,k-t+1} \nu_{s,k-t+1} \nonumber \\[1.5\jot]
	&= B_{s,k} \boldsymbol{\nu}_{s,k-\eta_{s,k}:k}
\end{align} 
where $\eta_{s,k}$ denotes the time elapsed since the last successful delivery in the $s$th link when we are at time $k$, and $B_{s,k}$ is a matrix of proper dimension. By (\ref{eq:lem6:et-dyn}) and the definitions of $\imath_{s,k}$, we have $\imath_{s,k} =  A_{s,k} B_{s,k} \E [  \boldsymbol{\nu}_{s,k-\eta_{s,k}:k} | \mathcal{I}^{d_s}_k, u_{s,k} \gamma_{s,k} = 0 ] =  A_{s,k} B_{s,k} \E [  \boldsymbol{\nu}_{s,k-\eta_{s,k}:k} | \mathcal{I}^{d_s}_{k-\eta_{s,k}}, u_{s, k-\eta_{s,k}} \gamma_{s, k-\eta_{s,k}} = 0, \dots, u_{s,k} \gamma_{s,k} = 0]$. Equivalently, we can write $\imath_{s,k} =  A_{s,k} B_{s,k} \E [  \boldsymbol{\nu}_{s,k-\eta_{s,k}:k} | u_{s, k-\eta_{s,k}} \gamma_{s, k-\eta_{s,k}} = 0, \dots, u_{s,k} \gamma_{s,k} = 0]$ since $\boldsymbol{\nu}_{s,k - \eta_{s,k}:k}$ is independent of $\mathcal{I}^{d_s}_{k-\eta_{s,k}}$. Moreover, note that $\Prob(\boldsymbol{\nu}_{s,k-\eta_{s,k}:k} | u_{s, k-\eta_{s,k}} \gamma_{s, k-\eta_{s,k}} = 0, \dots, u_{s,k} \gamma_{s,k} = 0)$ can be written as a linear combination of $\Prob(\boldsymbol{\nu}_{s,k-\eta_{s,k}:k} | \mathbf{u}_{s, k-\eta_{s,k}:k})$ for different values of $\mathbf{u}_{s, k-\eta_{s,k}:k}$ as $\boldsymbol{\nu}_{s,k-\eta_{s,k}:k}$ is independent of $\boldsymbol{\gamma}_{s, k-\eta_{s,k}:k}$.

We already showed that $\Prob(\boldsymbol{\nu}_{s,k-\eta_{s,k}:k} | \mathbf{u}_{s,k-\eta_{s,k}:k})$ is a symmetric distribution with respect to $\boldsymbol{\nu}_{s,k-\eta_{s,k}:k}$. Hence, $\Prob(\boldsymbol{\nu}_{s,k-\eta_{s,k}:k} | u_{s, k-\eta_{s,k}} \gamma_{s, k-\eta_{s,k}} = 0, \dots, u_{s,k} \gamma_{s,k} = 0)$ is a symmetric distribution with respect to $\boldsymbol{\nu}_{s,k-\eta_{s,k}:k}$. This implies that $\imath_{s,k} = 0$.
\end{IEEEproof}

The next lemma provides an equivalent loss function.
\begin{lemma}\label{lem:equiv-loss}\emph{
Let $\tilde{e}_{s,0}$ satisfy $K_{s,0} \nu_{s,0}$ and $\delta$ be specified based on $\E[x_{s,k} | \mathcal{I}^{d_s}_k]$ for $s \in \mathcal{M}$. Optimizing the loss function $\Phi(\epsilon,\delta)$ over $\epsilon \in \mathcal{E}$ is equivalent to optimizing the loss function $\Omega^{N}_{\epsilon} = \sum_{s \in \mathcal{M}} \Omega^{s,N}_{\epsilon}$ over $\epsilon \in \mathcal{E}$, where 
\begin{align}\label{eq:lemma:PSI}
	&\Omega^{s,N}_{\epsilon}(\tilde{e}_{s,0}) = \sum_{k=0}^{N} \Big\{ \theta_{s,k} \Prob_{\epsilon}(\mathbf{h}_{s,0:k-1} = 0) \nonumber\\[0.5\jot]
	& \times \E_{\epsilon} \Big[ u_{s,k} \big| \mathbf{h}_{s,0:k-1} = 0 \Big]\nonumber\\[0.5\jot]
	& + \Prob_{\epsilon}(\mathbf{h}_{s,0:k-1} = 0) \omega_{s,k} \E_{\epsilon} \Big[ \tilde{e}_{s,k}^T \tilde{e}_{s,k} \big| \mathbf{h}_{s,0:k-1} = 0 \Big] \nonumber\\[1.75\jot]	
	& + \Prob_{\epsilon}(\mathbf{h}_{s,0:k-1} = 0, h_{s,k} = 1) \nonumber \\[1.5\jot]
	&\times \E_{\epsilon} \Big[ \Omega^{s,k+1,N}_{\epsilon}(\tilde{e}_{s,k+1}) \big| \mathbf{h}_{s,0:k-1} = 0, h_{s,k} = 1 \Big] \Big\}
\end{align}
with $h_{s,k} = u_{s,k} \gamma_{s,k}$ and
\begin{align*}
	\Omega^{s,k,N}_{\epsilon}(\tilde{e}_{s,k}) = \sum_{t=k}^{N} \E \Big[ \theta_{s,t} u_{s,t} + \omega_{s,t} \tilde{e}_{s,t}^T \tilde{e}_{s,t} \Big]
\end{align*}
for $k \in \mathbb{N}_{[1,N]}$ when $\tilde{e}_{s,k}$ satisfies $K_{s,k} \nu_{s,k}$.	}
\end{lemma}

\begin{IEEEproof}
From the definition of the loss function $\Phi(\epsilon,\delta)$ when $\delta$ is chosen based on $\hat{x}_{s,k} = \E[x_{s,k} | \mathcal{I}^{d_s}_k]$, we can write
\begin{align*}
	\Phi(\epsilon,\delta) &= \textstyle \sum_{k=0}^{N} \sum_{s \in \mathcal{M}} \E \Big[ \theta_{s,k} u_{s,k} + \omega_{s,k} \hat{e}_{s,k}^T \hat{e}_{s,k} \Big]\\[1.5\jot]
	&= \textstyle \sum_{k=0}^{N} \sum_{s \in \mathcal{M}} \E \Big[ \theta_{s,k} u_{s,k} + \omega_{s,k} \tilde{e}_{s,k}^T \tilde{e}_{s,k} + \tr Q_{s,k} \Big]
\end{align*}
where in the second equality we used the tower property of conditional expectations. Define the loss function $\Omega^{N}_{\epsilon} := \sum_{k=0}^{N} \sum_{s \in \mathcal{M}} \E [ \theta_{s,k} u_{s,k} + \omega_{s,k} \tilde{e}_{s,k}^T \tilde{e}_{s,k} ]$. Following the fact that $\tr Q_{s,k}$ is independent of $\epsilon$, to optimize  $\Phi(\epsilon,\delta)$ over $\epsilon \in \mathcal{E}$, it suffices to optimize $\Omega^{N}_{\epsilon}$ over $\epsilon \in \mathcal{E}$. In addition, observe that from the law of total probability, we have
\begin{align}\label{eq:identity2}
	&\Prob_{\epsilon}(h_{s,0} = 1) + \Prob_{\epsilon}(\mathbf{h}_{s,0:t} = 0) \nonumber\\[1.75\jot]
	&\textstyle  \qquad \qquad  + \sum_{t'=1}^{t} \Prob_{\epsilon}(\mathbf{h}_{s,0:t'-1} = 0, h_{s,t'} = 1) = 1
\end{align}
for any $t \in \mathbb{N}_{[0,N]}$. Applying the law of total expectation for the terms $\E[ \theta_{s,k} u_{s,k}]$ and $\E[\omega_{s,k}  \tilde{e}_{s,k}^T \tilde{e}_{s,k}]$ in $\Omega^{s,N}_{\epsilon}$ on a partition provided by the identity (\ref{eq:identity2}) for $t = k-1$, repeating this procedure for $k \in \mathbb{N}_{[1,N]}$, and using the definition of $\Omega^{s,k,N}_{\epsilon}$, we can obtain (\ref{eq:lemma:PSI}).
\end{IEEEproof}

The next two lemmas are related to symmetric decreasing rearrangements of non-negative functions. For the proofs of these lemmas, see, e.g., \cite{brock2000} and \cite{alvino1991}.

\begin{lemma}\label{lemma:GHL}\emph{
Let $f$ and $g$ be non-negative functions defined on $\mathbb{R}^n$ that vanish at infinity. Then,
\begin{align}
	\int_{\mathbb{R}^n} f(x) g(x) dx \leq \int_{\mathbb{R}^n} f^*(x) g^*(x) dx.
\end{align}
}
\end{lemma}
\vspace{2mm}

\begin{lemma}\label{lemma:major}\emph{
Let $\mathcal{B}(r) \subseteq \mathbb{R}^n$ be a ball of radius $r$ centered at the origin, and $f$ and $g$ be non-negative functions defined on $\mathbb{R}^n$ that vanish at infinity and satisfy
\begin{align}
	\int_{\mathcal{B}(r)} f^*(x) dx \leq \int_{\mathcal{B}(r)}  g^*(x) dx 
\end{align}
for all $r \geq 0$. Then,
\begin{align}
	\int_{\mathcal{B}(r)} h(x) f^*(x) dx \leq \int_{\mathcal{B}(r)} h(x) g^*(x) dx
\end{align}
for all $r \geq 0$ and any non-negative non-increasing function $h$.}
\end{lemma}

\bibliography{../../mybib}
\bibliographystyle{ieeetr}
 
\vspace{-0.5cm}
\begin{IEEEbiographynophoto}{Touraj~Soleymani} received his B.S. and M.S. degrees both in aeronautical engineering from Sharif University of Technology, Iran, in 2008 and 2011, respectively, and his Ph.D. degree in electrical and computer engineering from the Technical University of Munich, Germany, in 2019. He is currently a research associate at the Department of Electrical and Electronic Engineering, Imperial College London, United Kingdom. He was a research associate at the School of Electrical Engineering and Computer Science, Royal Institute of Technology, Sweden, from 2019 to 2022; and was a research scholar at the Institute of Artificial Intelligence, University of Brussels, Belgium, from 2012 to 2014, at the Institute for Advanced Study, Technical University of Munich, Germany, from 2014 to 2017, and at the School of Electrical Engineering and Computer Science, Royal Institute of Technology, Sweden, from 2017 to 2019. His research interests include control, learning, communication, optimization, game theory, multi-agent systems, cyber-physical systems, and swarm robotics systems. He received the Outstanding Student Award from Sharif University of Technology in 2011, the Best Paper Award at the International Conference on Intelligent Autonomous Systems in 2014, and the Best Paper Award Finalist at the International Workshop on Discrete Event Systems in 2018.
\end{IEEEbiographynophoto}
\vfill

\newpage

\begin{IEEEbiographynophoto}{Deniz~G\"{u}nd\"{u}z} received the B.S. degree in electrical and electronics engineering from Middle East Technical University (METU), Turkey, in 2002, and the M.S. and Ph.D. degrees in electrical engineering from the NYU Tandon School of Engineering (formerly Polytechnic University), USA, in 2004 and 2007, respectively. After his Ph.D. degree, he was a Post-Doctoral Research Associate at Princeton University, a Consulting Assistant Professor at Stanford University, and a Research Associate at CTTC, Barcelona, Spain. He also held visiting positions at the University of Padova from 2018 to 2020 and Princeton University from 2009 to 2012. In September 2012, he joined the Department of Electrical and Electronic Engineering, Imperial College London, UK, where he is currently a Professor of information processing. He also serves as the Deputy Head of the Intelligent Systems and Networks Group and a part-time Faculty Member at the University of Modena and Reggio Emilia, Italy. His research interests include communications and information theory, machine learning, and privacy. He is a Distinguished Lecturer of the IEEE Information Theory Society from 2020 to 2022. He was a recipient of the Consolidator in 2022 and Starting in 2016 Grants of the European Research Council (ERC), the IEEE Communications Society-Communication Theory Technical Committee (CTTC) Early Achievement Award in 2017, and several best paper awards. He is an Area Editor of the IEEE Transactions on Information Theory, the IEEE Transactions on Communications, and the IEEE Journal on Selected Areas in Communications (JSAC), Special Series on Machine Learning in Communications and Networks. He also serves as an Editor for the IEEE Transactions on Wireless Communications.
\end{IEEEbiographynophoto}
\vfill

\end{document}